\documentclass[preprint]{aastex62}
\usepackage{comment}
\usepackage{amsmath, float}
\usepackage[utf8x]{inputenc}
\usepackage{graphicx}
\usepackage{amstext}
\usepackage{subfigure}
\usepackage{rotating}
\usepackage{threeparttable}
\usepackage{longtable}

\graphicspath{{./}{figures/}}

\received{23-Feb-2021}
\revised{--}
\accepted{09-Apr-2021}

\submitjournal{ApJ}

\begin{document}

\title{Thermal desorption of astrophysical relevant ice mixtures of acetaldehyde and acetonitrile from olivine dust\footnote{Released on --}}

\correspondingauthor{John Robert Brucato}
\email{john.brucato@inaf.it}
 
 \author[0000-0002-0786-7307]{Maria Angela Corazzi}
 \affil{Department of Physics and Astronomy, University of Florence\\
Via G. Sansone 1\\
Sesto Fiorentino (Firenze), 50019, Italy}
\affil{INAF - Astrophysical Observatory of Arcetri\\
 Largo E. Fermi 5\\
 Firenze, 50125, Italy}
\author{John Robert Brucato}
\affil{INAF - Astrophysical Observatory of Arcetri\\
 Largo E. Fermi 5\\
 Firenze, 50125, Italy}
\author{Giovanni Poggiali}
\affil{Department of Physics and Astronomy, University of Florence\\
Via G. Sansone 1\\
Sesto Fiorentino (Firenze), 50019, Italy}
\affil{INAF - Astrophysical Observatory of Arcetri\\
 Largo E. Fermi 5\\
 Firenze, 50125, Italy}
\author{Linda Podio}
\affil{INAF - Astrophysical Observatory of Arcetri\\
 Largo E. Fermi 5\\
 Firenze, 50125, Italy}
\author{Davide Fedele}
\affil{INAF - Astrophysical Observatory of Torino\\
 Via Osservatorio 20\\
Pino Torinese, 10025, Italy}
\affil{INAF - Astrophysical Observatory of Arcetri\\
 Largo E. Fermi 5\\
 Firenze, 50125, Italy}
\author{Claudio Codella}
\affil{INAF - Astrophysical Observatory of Arcetri\\
 Largo E. Fermi 5\\
 Firenze, 50125, Italy}

\begin{abstract}
Millimeter and centimeter observations are discovering an increasing number of interstellar complex organic molecules (iCOMs) in a large variety of star forming sites, from the earliest stages of star formation to protoplanetary disks and in comets. In this context it is pivotal to understand how the solid phase interactions between iCOMs and grain surfaces influence the thermal desorption process and, therefore, the presence of molecular species in the gas phase. In laboratory, it is possible to simulate the thermal desorption process deriving important parameters such as the desorption temperatures and energies. 
We report new laboratory results on temperature-programmed desorption (TPD) from olivine dust of astrophysical relevant ice mixtures of water, acetonitrile, and acetaldehyde. We found that in the presence of grains, only a fraction of acetaldehyde and acetonitrile desorbs at about 100 K and 120 K respectively, while 40$\%$ of the molecules are retained by fluffy grains of the order of 100 $\mu$m up to temperatures of 190 - 210 K. 
In contrast with the typical assumption that all molecules are desorbed in regions with temperatures higher than 100 K, this result implies that about 40$\%$ of the molecules can survive on the grains enabling the delivery of volatiles towards regions with temperatures as high as 200 K and shifting inwards the position of the snowlines in protoplanetary disks. These studies offer a necessary support to interpret observational data and may help our understanding of iCOMs formation providing an estimate of the fraction of molecules released at various temperatures. 
\end{abstract}

\keywords{acetonitrile --- acetaldehyde --- analog ice mixtures --- prebiotic molecules --- interstellar complex organic molecules (iCOMs) --- Temperature Programmed Desorption (TPD) curves --- protoplanetary disks --- grain surface --- space chemistry---methods: laboratory}

\section{Introduction} 
Millimeter and centimeter observations are discovering an increasing number of interstellar complex organic molecules (iCOMs) in a large variety of star forming sites.  iCOMs are defined as chemical compounds with at least six atoms in which at least one is carbon (e.g., \citealt{Blake1987}, \citealt{Herbst2009}). The advent of large interferometers in the (sub-)mm range, such as IRAM-NOEMA and ALMA, have shown the presence of iCOMs from the earliest stages of star formation (see e.g., the review by \citealt{Caselli2012}): i.e. in pre-stellar dense cores (e.g., \citealt{Bacmann2012}, \citealt{Vastel2014}), in hot corinos around protostars (e.g., \citealt{Cazaux2003}, \citealt{Ceccarelli2007}), 
and in the associated jets and outflows (e.g., \citealt{bachiller1997}, \citealt{Arce2008}, \citealt{Codella2010}, \citealt{Lefloch2017}). 
In the case of protoplanetary disks, complex species are hardly detected because the region where the dust temperature is high enough to let water ice and iCOMs to thermally desorb (T$>$100 K) is very small ($\lesssim$ 5 au) for Solar-like stars (e.g., \citealt{Cieza2016}). Only methanol (\citealt{walsh2016}, \citealt{Podio2020a}), acetonitrile (e.g.,
\citealt{Oberg2015}), and formic acid \citep{Favre2018} have been observed
so far. A new fascinating perspective was recently provided by FU Ori objects in which the young central star undergoes a strong accretion burst, hence a sudden increase in brightness which leads to heating of the surrounding disk and to a quick expansion of the molecular snow lines to larger radii. This phenomenon was first observed in V883 Ori by \cite{Cieza2016}, where five iCOMs thermally desorbed from the disk were detected out to radii of $\sim$ 160 au: first methanol \citep{vanthof2018}, and then acetone, acetonitrile, acetaldehyde, and methyl formate \citep{Lee2019}. \\
To correctly interpret the distribution and abundances of iCOMs observed along the formation process of a Sun-like stars, we need to first comprehend their formation processes and the mechanisms responsible for their release in gas-phase. Thermal desorption can be characterised through laboratory experiments using interstellar ice analogs deposited on grains which are similar to interstellar ones.\\
The laboratory studies reported in this paper are triggered in the general context of understanding how the solid phase interactions between molecules and grain surfaces can significantly influence the thermal desorption process and so the presence of molecular species in the gas phase. In laboratory, it is possible to simulate the thermal desorption process of iCOMs through Temperature Programmed Desorption (TPD) analysis, deriving important parameters such as the thermal desorption temperatures and energies.
Up to now, TPD experiments have been carried out mainly from graphite and amorphous water ice surfaces (e.g., \citealt{Collings2004}, \citealt{Hama2011}, \citealt{Shi2015}, \citealt{Chaabouni2017}). Therefore as far as we know, TPD experiments from grain surfaces are lacking in the literature, but grain-molecule interaction is a fundamental aspect: mineral matrices can selectively adsorb, protect, and allow the iCOMS concentration on their surface. Molecules can interact on the mineral surface through Van der Waals-like forces and dipole-dipole interactions as it has been routinely demonstrated through vibrational spectroscopic methods such as infrared and Raman spectroscopy \citep{Jacob2018}. Moreover, molecules can diffuse inside the grains when the submicron interstellar grains begin to accrete into hundreds of microns fluffy dust. The presence of grains can therefore influence the desorption and release of the iCOMs in the gas phase and therefore, the molecule-grain interaction cannot be neglected in thermal desorption studies. \\
We performed for the first time TPD experiments of acetonitrile (CH$_{3}$CN) and acetaldehyde (CH$_{3}$COH) both pure and mixed with water from micrometric grains of silicate olivine ((Mg,Fe)$_{2}$SiO$_{4}$) used as dust analog on which the icy mixtures were condensed at 17 K.
In Section 2, we present the astrophysical importance of acetonitrile, acetaldehyde, and the choice of olivine as relevant cosmic dust analogs. In Section 3, we show the experimental set-up, method, and procedure. In Section 4, the results of TPD analysis of acetaldehyde and acetonitrile mixtures, discussing the case of ice mixtures desorbed from the olivine substrate, are reported. In Section 5, we discuss the obtained results in the context of astrophysical observations of acetaldehyde and acetonitrile. Finally, we summarise our conclusions in Section 6.

\section{Choice of molecules and olivine mineral}
\textbf{Acetonitrile} (CH$_{3}$CN, molecular weight 41 atomic mass units a.m.u.) was found in many regions both within and outside the Solar System: in the Titan atmosphere (e.g., \citealt{Cordiner2015}, \citealt{Iino2020},   \citealt{Thelen2019}), in the comet 67P / Churyumov-Gerasimenko \citep{Goesmann2015}, in the cometary coma of Hale - Bopp \citep{Woodney2002}), in the molecular cloud Sgr B$_{2}$(N) (e.g., \citealt{Willis2020}), in the high and low mass protostars (e.g., \citealt{Taniguchi2020}, \citealt{Andron2018}), in hot cores (e.g., \citealt{Bogelund2019}), in the protoplanetary disks (e.g., \citealt{Oberg2015}, \citealt{Bergner2018}) and it is among the five iCOMs desorbed by the frozen mantle of the circumstellar disk around the V883 Ori protostar \citep{Lee2019}. Moreover, it is among the most commonly detected organic molecules in disks \citep{Gal2019}.
To explain the large abundance observed in the gas phase of acetonitrile, many works invoked chemical reactions occurring on the surface of the grains (\citealt{Oberg2015}, \citealt{Loomis2018}).\\
\textbf{Acetaldehyde} (CH$_{3}$COH, molecular weight 44 a.m.u.) is one of the five complex molecules that are thermally desorbed by the FU Ori system V883 Ori \citep{Lee2019}. It was also observed in the interstellar medium, in low- and high-mass protostars (e.g., \citealt{Bianchi2019}, \citealt{Guzman2018}), in hot corinos (e.g., \citealt{Taquet2015}, \citealt{Codella2016}) and comets (e.g., \citealt{Biver2019}). 
The reaction pathways involving the iCOMs described by the C$_{2}$H$_{n}$O formula, such as acetaldehyde (CH$_{3}$CHO), are still matter of debate \citep{Chuang2020}. Many works show how acetaldehyde can be synthesized directly on the grain surfaces (e.g., \citealt{Garrod2006}, \citealt{Oberg2009}) starting from HCO and CH$_{3}$ radicals on CO-rich ices \citep{Lamberts2019} or through surface chemistry at 10 K on C$_{2}$H$_{2}$ ices with H-atoms and OH-radicals \citep{Chuang2020} or through UV photoprocessing of interstellar ice analogs \citep{MartinD2020} or through 5 keV electron irradiation of CO and CH$_{4}$ \citep{Bennett2005a} and of CO$_{2}$ and C$_{2}$H$_{4}$ ice mixtures \citep{Bennett2005b}. Moreover, \cite{Hudson2008} showed that acetaldehyde is synthesized by ion-irradiation of nitrile-containing ices at 10K. Other works claimed that it is not clear whether acetaldehyde is formed on the icy surfaces of interstellar grains or through gas phase reactions \citep{Enrique2016}. Therefore, the laboratory study of the acetonitrile and acetaldehyde desorption and their interaction with dust grains is fundamental.\\
\\
Silicates are ubiquitous in space, from comets observed from both ground observations (e.g., \citealt{Shinnaka2018} and \citealt{Ootsubo2020} with Subaru 8.2 m telescopes or \citealt{Picazzio2019} with 4.1m SOAR telescopes) and from space missions (e.g., \citealt{Brownlee2006} for Stardust Comet Sample Return Mission and \citealt {BockeleeMorvan2017} for Rosetta Mission) to protoplanetary disks (e.g., \citealt{Cohen1985}, \citealt{Przygodda2003}, \citealt{vanboekel}, \citealt{Natta2007}, \citealt{henning2010}).
One of the most common silicate grain in space is \textbf{olivine} ((Mg,Fe)$_{2}$SiO$_{4}$), an isomorphic mixture of forsterite (Mg$_{2}$SiO$_{4}$) and fayalite (Fe$_{2}$SiO$_{4}$). In our experiment, we used olivine made up of 80$\%$ of forsterite and 20$\%$ of fayalite. Fayalite (Fe$_{2}$Si0$_{4}$), the iron end-member of the olivine group, was found in the interstellar medium and meteorite \citep{Boruah2017}. Also forsterite Mg$_{2}$SiO$_{4}$, the magnesium- rich end-member of the olivine group, was found in different astrophysical environments from meteorites \citep{Weinbruch2000} to protoplanetary disk \citep{Fujiyoshi2015}. 
Moreover, through IRAS low-resolution spectra, olivine dust was detected as common material in the circumstellar disks or/and shells of Herbig Ae/Be stars (e.g., \citealt{Chen2000}) and around T Tauri stars (e.g., \citealt{Honda2003}). 
More recently, the mid-infrared spectrum obtained through Spitzer Space Telescope revealed the presence of crystalline silicate in a cold, infalling, protostellar envelope of the Orion A protostar HOPS-68 \citep{Poteet2011}.
Protostar EX Lupi is the prototype of EXORs like objects, that are similar to FUORs objects mentioned above but their outbursts are shorter and recursive in time, and it underwent its most important explosion in 2008, when its brightness increased by a factor of 30 for six months, due to the high accretion from the circumstellar disk on the star. \cite{Abraham2020} observed the system during the explosion and discovered the crystallization of the amorphous silicate grains due to the heating of the disk. In particular, they observed silicates of the order of micrometer size.

\section{Experimental method and procedure}
\subsection{Experimental setup and thermal desorption diagnostic} 
We assembled an ultra-high vacuum (UHV) chamber (P $\sim$ $6.68\cdot 10^{-10}$ mbar) with feed-throughs for gas phase deposition from a pre-chamber (P $\sim 10^{-7}$ mbar). 
The UHV chamber was equipped with the Hiden Analytical 3F RC 301 Pic Quadrupole Mass Spectrometer (HAL 3F RC) for mass spectrometry. The ion source is an electron impact ionizer with twin-oxide-coated iridium filaments and the detector is a pulse ion counting single channel electron multiplier, which allows us to analyze masses from 1 to 300 atomic mass units (a.m.u.). 
The chamber interfaces with ARS closed cycle helium cryocooler able to get temperature of 17 K.
\subsection{Samples preparation} 
\begin{figure}
\begin{center}
\includegraphics[scale=0.38]{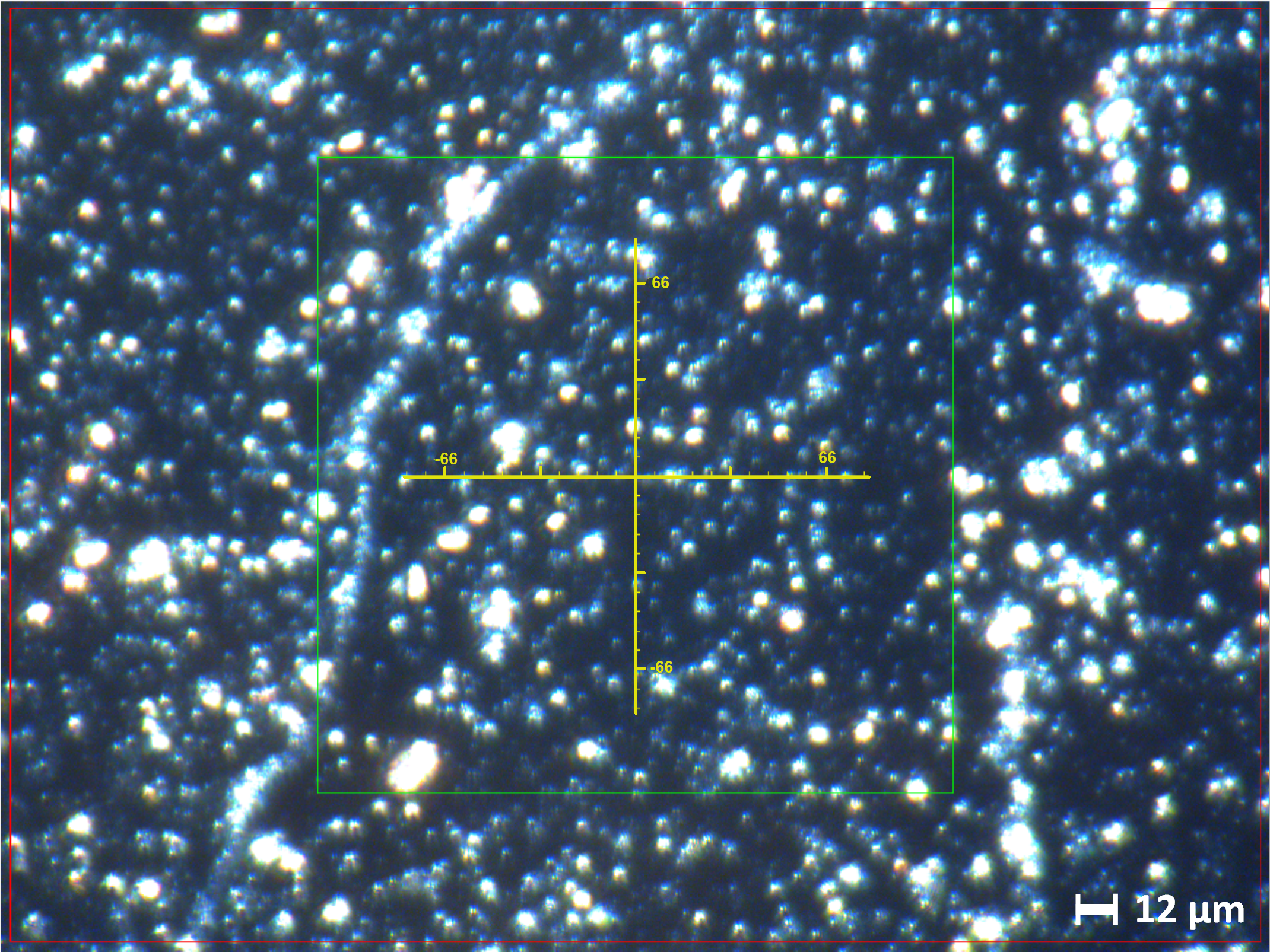}
\caption{Olivine grains smaller than 5$\mu m$. The image was obtained with the Hyperion microscope.}\label{graniolivina}
\end{center}
\end{figure}

The CH$_{3}$COH and CH$_{3}$CN molecules were purchased from Sigma Aldrich, Merck corporation with a purity 99.5$\%$.
Pure H$_{2}$O, CH$_{3}$CN, CH$_{3}$COH, and mixtures CH$_{3}$CN:H$_{2}$O (1:2), CH$_{3}$COH:H$_{2}$O (1:2), CH$_{3}$CN:CH$_{3}$COH (1:6), and CH$_{3}$CN:CH$_{3}$COH:H$_{2}$O (1:1:3) were prepared. To obtain the gas mixtures with the described proportions, the gas mixing was controlled by their partial pressures inside the pre-chamber.\\
We first studied the thermal desorption of these samples condensed directly on the cold finger of the cryostat at 17 K made with a smooth nickel-plate. Then to simulate a process that can realistically take place in hot star-forming regions, we carried out the thermal desorption experiments of pure molecules and mixtures condensed on micrometric grains of olivine. So, the cold finger was covered with hundreds micron thick layer of olivine grains with size smaller than 5 $\mu m$ used as a substrate.
To obtain in the laboratory substrate of suitable dimensions, terrestrial bulk olivine was first ground using a planetary mill Retsch PM 100. In this way, we obtained dust of different grain sizes. To select micrometric grains, two procedures were adopted: dry sieving to select grains with size d$<$20 $\mu m$ and then a methanol sedimentation procedure to select grains with even smaller dimensions. The use of methanol as a solvent is justified by its high volatility, but also because it is a polar solvent able to remove organic contamination from the natural mineral. With sedimentation, grains with dimensions d$<$5 $\mu m$ were selected. The grain size measurement was carried out with the Bruker Hyperion 1000 microscope. 
Figure \ref{graniolivina} shows the image of micrometric olivine dust.\\

\subsection{TPD experiments and analysis} Once the gas mixture of iCOMs was prepared in the pre-chamber following the partial pressures, it was deposited through a valve system inside the UHV chamber on the cold finger of the cryostat cooled to 17 K. The mixture condensed on the cold finger (clean or covered with the olivine substrate). In all measurements during the gas mixtures deposition, we observed that the pressure in the UHV chamber increased from $6.68\cdot 10^{-10}$ mbar without ever exceeding a maximum pressure of $10^{-8}$ mbar. The pressure maintained this maximum value for about 1 second. The surface exposure to a gas is generally measured by the Langmuir unit (L). Assuming a sticking coefficient equal to 1, i.e. each gas molecule that reaches the surface sticks to it, 1 L ($1L=10^{-6}$torr$\cdot$ sec) results in the formation of a monolayer of gas molecules adsorbed on the surface. Since during the deposit the maximum pressure in the UHV chamber was $10^{-8}$ mbar for one second, we could say that we were in a regime of 0.01 L, so we were in a substrate regime. In the following sections, we report the amounts of the molecules adsorbed on the surface in partial pressure instead of L unit since the first one is a direct measure in our experimental apparatus.
After condensation, the sample was heated at a constant rate of 1.21 K$\cdot$ sec$^{-1}$. As the cold finger warmed, the condensed molecules desorbed, entered the mass spectrometer, and were detected. We followed the signal detected by the mass spectrometer as the temperature increased for selected masses from 1 to 300 a.m.u. and so, the temperature programmed desorption (TPD) curves were obtained.\\
The desorption temperature (T$_{d}$) and the desorption energy (E$_{d}$) were measured by fitting the TPD curves with the Polanyi-Wigner equation (e.g., \citealt{attard1998}) with a pre-exponential factor of 10$^{12}$ sec$^{-1}$, found as best fit value for both acetaldehyde and acetonitrile. A detailed description of the TPD analysis and of the determination of the desorption temperature and desorption energy was reported in our previous work \cite{Agnola2020}. 

\section{Thermal desorption from olivine grains results}
\begin{figure*}
\begin{center}
{\includegraphics[width=14cm]{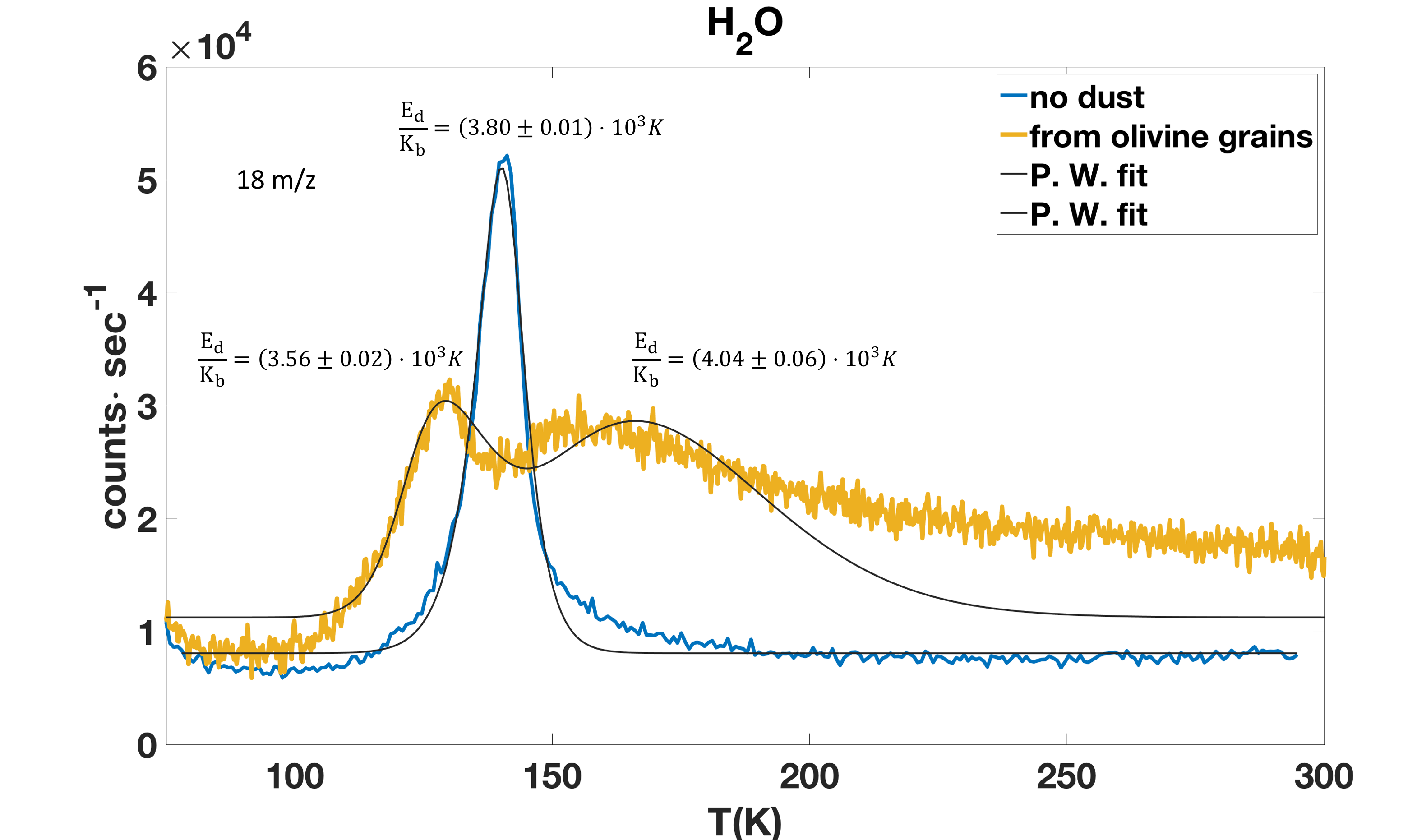}}
\caption{The figure shows two experiments. In both experiments we deposited in the UHV chamber the same amount of water (partial pressure of water in the pre--chamber 0.68 mbar). \textbf{Blue curve}: water desorption directly from the cold finger of cryostat cooled to 17 K. \textbf{Yellow curve}: water desorption from olivine dust. In presence of olivine, we found a first desorption peak lower than that found in the absence of grains and then we found a second wider desorption peak. In presence of crystalline grains, the molecules  did not desorb all together at T$_{des}$: some of them desorbed but others remained tied to the grains and desorbed with a delay at a higher temperature. This second desorption shows the interactions between the molecules and the grains.
\textbf{Black lines}: the TPD curves were fitted through the Polanyi - Wigner equation}\label{confrontoacqua}
\end{center}
\end{figure*}

\begin{figure*}
\begin{center}
{\includegraphics[width=14cm]{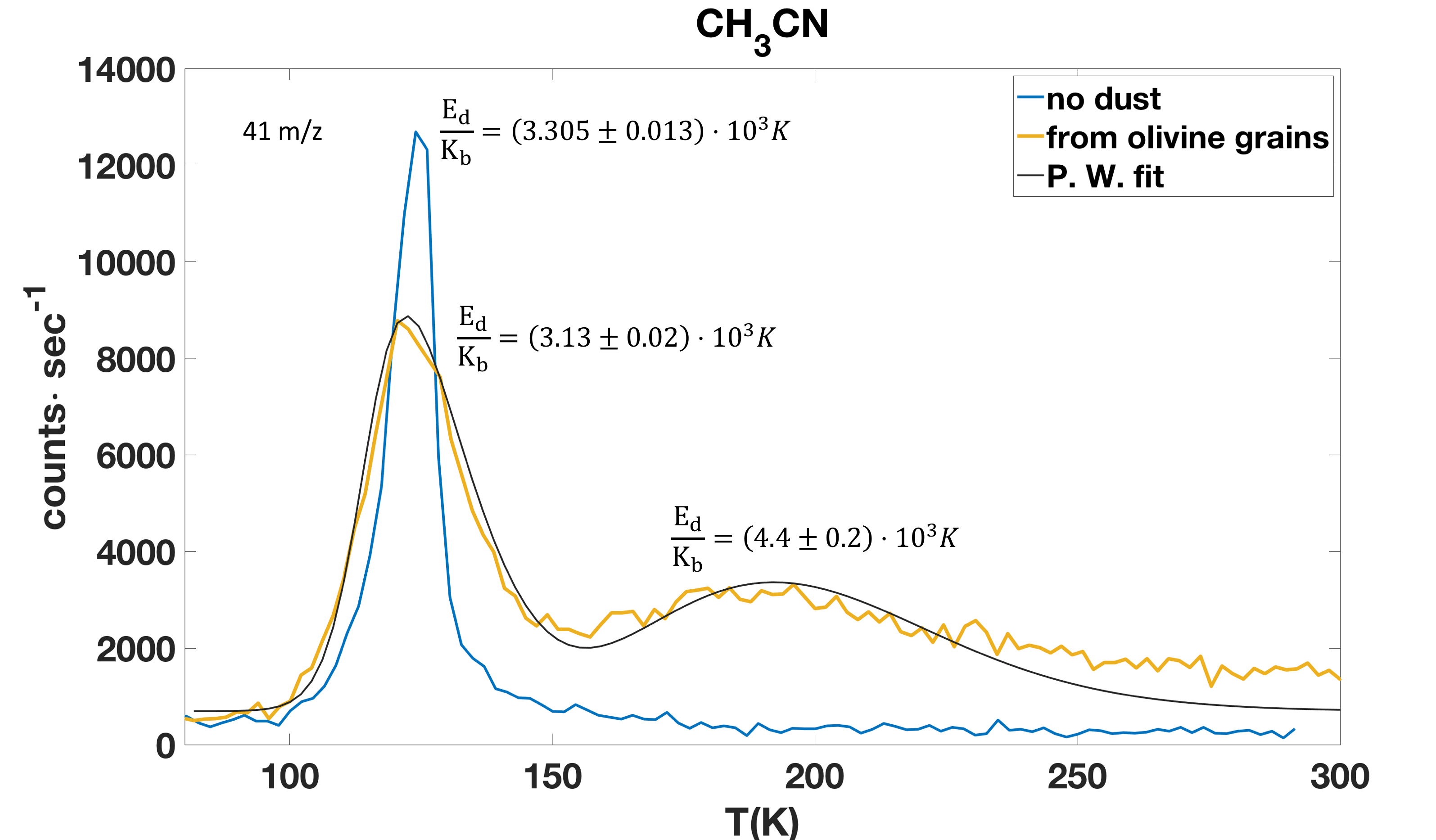}}
\caption{The figure shows two experiments. In both experiments we deposited in the UHV chamber the same amount of acetonitrile CH$_{3}$CN (partial pressure of CH$_{3}$CN in the pre--chamber 1 mbar). \textbf{Blue \textbf{curve}}: CH$_{3}$CN desorption directly from the cold finger of cryostat cooled to 17 K. \textbf{Yellow \textbf{curve}}: CH$_{3}$CN desorption from olivine dust. In presence of olivine, we found a first desorption peak lower in intensity than that found in the absence of grains and then we found a second wider deosrption peak, that shows the interactions between the molecules and the grains.
\textbf{Black line}: the double desorption was fitted through the Polanyi - Wigner equation}\label{confrontoacetonitrile}
\end{center}
\end{figure*}

In this section, we report the results obtained with the seven samples listed above.
Figures \ref{} and \ref{confrontoacetonitrile} show in detail the TPD curves of pure H$_{2}$O (18 m/z) and pure CH$_{3}$CN (41 m/z) desorbed both from the cold finger of the cryostat (blue curves) and from micrometric olivine grains (yellow curves).
As visible in Figure \ref{confrontoacqua} when the molecules desorbed directly from the cold finger of the cryostat, only one desorption peak was obtained. On the other hand, when the same number of molecules (i.e. same partial pressures) condensed on the micrometric grains of olivine, two different desorption peaks appeared. Specifically, the first desorption peak was found at temperature lower than that found in the absence of grains i.e. when desorption occurred from the smooth nickel-plated cold finger. Water ice TPD measurement (blue curve in Figure \ref{confrontoacqua}) shows that as the temperature increased, the water molecules began to desorb at around 110 K until reaching the desorption peak at 141.2 K with the counts detected by the mass spectrometer increasing from $6.6\cdot 10^{3}$ to $5.2\cdot 10^{4}$ counts$\cdot$sec$^{-1}$. At 165 K all the water molecules deposited were desorbed and as the temperature increased further the counts decreased until they returned to the initial value. \\
In the presence of the olivine substrate (Figure \ref{confrontoacqua}, yellow curve), we observed that 50$\%$ of the water molecules desorbed at 128.5 K where the signal at 18 m/z increased to $3.2\cdot 10^{4}$ counts$\cdot$sec$^{-1}$. The rest of water ice desorbed in a wider temperature range which shows a peak at 166.4 K. If in the absence of grains at 160 K the mass counts returned to the initial value, in presence of grains the counts were still higher than the background with a value of $2.2\cdot 10^{4}$ counts$\cdot$ sec$^{-1}$ at 207 K and $1.6\cdot 10^{4}$ counts$\cdot$sec$^{-1}$ still at 300 K, showing a gradual and prolonged desorption of water molecules from the dust grains. This result reveals that the water remains trapped by the olivine grains showing a desorption drift.\\ 
We observed a similar desorption behavior for CH$_{3}$CN. In this case the partial pressure of the pre-chamber was 1 mbar. Figure \ref{confrontoacetonitrile} shows that as the temperature increased, the CH$_{3}$CN molecules started to desorb from the cold finger at around 100 K reaching the maximum peak at 124.0 K where the mass counts got the value of $1.27\cdot 10^{4}$ counts$\cdot$sec$^{-1}$. At higher temperature the counts decreased and returned to the initial value of the background already at 150 K. TPD curve shows a single desorption peak at 124.0 K where all the CH$_{3}$CN molecules deposited were desorbed from the cold finger of the cryostat and passed into the gas phase. \\
Depositing the same amount of CH$_{3}$CN onto the olivine dust, the mass counts of the first peak of the yellow curve increased to $3.2\cdot 10^{4}$ counts$\cdot$sec$^{-1}$ and 30$\%$ of molecules desorbed with a delay at 190.5 K. The yellow curve showed a second desorption peak of intensity $3\cdot 10^{3}$counts$\cdot$sec$^{-1}$. So in both cases, H$_{2}$O and CH$_{3}$CN presented a first desorption peak followed by a second one at higher desorption temperatures, 166.4 K for H$_{2}$O and 190.5 K for CH$_{3}$CN and hence with higher desorption energies. The Table \ref{tablefit} shows the best fit values for the desorption temperatures and energies obtained through the Polanyi - Wigner equation.\\
Furthermore, during the water TPD measurement, the first peak preceded the desorption peak found in the absence of grains by more than 10 K, while during the desorption of CH$_{3}$CN we did not observe this shift.

\begin{table*} \small
\begin{threeparttable}
\caption{\textbf{Best-fit values of TPD curves}}
\centering
\tiny
\begin{tabular}{p{0.25\textwidth}|p{0.1\textwidth}p{0.15\textwidth}|p{0.1\textwidth}p{0.15\textwidth}}
&\textbf{No dust}&&\textbf{Olivine grains}&\\
\cline{2-5}
\cline{2-5}
\textbf{Ice mixture ratios} &\textbf{T$_{d}$(K)} &\textbf{E$_{d}$/K$_{b}$(K)} &\textbf{T$_{d}$(K)} &\textbf{E$_{d}$/K$_{b}$(K)}  \\
\hline
\hline
Pure water &&\\
\textbf{18 m/z}&141.2&$(3.80 \pm 0.01)\cdot 10^{3}$&128.5&$(3.56 \pm 0.02)\cdot 10^{3}$\\   
&...&...&166.4&$(4.04\pm 0.06)\cdot 10^{3}$\\
\hline
\hline
CH$_{3}$CN&&\\
\textbf{41 m/z}&124.0&$(3.305\pm 0.013)\cdot 10^{3}$&124.6&$(3.13 \pm 0.02)\cdot 10^{3}$\\
&...&...&190.5&$(4.4\pm 0.2)\cdot 10^{3}$\\
\hline
\hline
CH$_{3}$COH&&\\
\textbf{44 m/z}&104.9&$(2.847 \pm 0.003)\cdot 10^{3}$&103.0&dissociated in HCO and CH$_{3}$\\
\cline{2-5}
15m/z&...&...&103.0&$(2.665\pm 0.014)\cdot 10^{3}$\\
&...&...&170.2&$(3.4 \pm 0.8)\cdot 10^{3}$\\
\cline{2-5}
29m/z&...&...&103.0&$(2.673\pm 0.010)\cdot 10^{3}$\\
&...&...&170.2&$(3.5 \pm 0.5)\cdot 10^{3}$\\
\hline
\hline
CH$_{3}$CN:H$_{2}$O=1:2&&\\
41m/z&121.4&$(3.257 \pm 0.007)\cdot 10^{3}$&126.2&$(3.36 \pm 0.03)\cdot 10^{3}$\\
&...&...&196.4&$(5.91\pm 0.15)\cdot 10^{3}$\\
\cline{2-5}
44 m/z&121.4&$(3.304\pm 0.005)\cdot 10^{3}$&124.0&$(3.34\pm 0.03)\cdot 10^{3}$\\
&...&...&193.0&$(4.80 \pm 0.07)\cdot 10^{3}$\\
\hline
\hline
CH$_{3}$COH:H$_{2}$O=1:2&&\\
44 m/z&107.5&$(3.079 \pm 0.011)\cdot 10^{3}$&108.2&dissociated\\
\hline
\hline
CH$_{3}$CN:CH$_{3}$COH=1:6&&\\
29 m/z&111.2&$(2.955 \pm 0.005)\cdot 10^{3}$&110.7&$(2.89 \pm 0.02)$\\
&...&...&149.4&$(3.1 \pm 0.7)\cdot 10^{3}$\\
\cline{2-5}
15 m/z&111.2&$(2.967 \pm 0.005)\cdot 10^{3}$&110.7&$(2.88 \pm 0.02)\cdot 10^{3}$\\
&...&...&149.4&$(3 \pm 1)\cdot 10^{3}$\\
\cline{2-5}
41 m/z&130.7&$(3.380 \pm 0.012)\cdot 10^{3}$&122.9&$(3.26 \pm 0.04)\cdot 10^{3}$\\
&...&...&190.2&$(4.4 \pm 0.3)\cdot 10^{3}$\\
\cline{2-5}
44 m/z&111.2-130.7&$(3.404 \pm 0.012)\cdot 10^{3}$&110.7-122.9&$(2.80 \pm 0.05)\cdot 10^{3}$\\
&...&...&190.3&$(4.3 \pm 0.2)\cdot 10^{3}$\\
\hline
\hline
CH$_{3}$CN:CH$_{3}$COH:H$_{2}$O=1:1:3&&\\
44m/z&121.9&$(3.275 \pm 0.018)\cdot 10^{3}$&123.0&$(3.33\pm 0.04)\cdot 10^{3}$\\
&...&...&206.6&$(4.7 \pm 0.3)\cdot 10^{3}$\\
\cline{2-5}
41 m/z&120.0&$(3.24 \pm 0.03)\cdot 10^{3}$&123.0&$(3.36 \pm 0.07)\cdot 10^{3}$\\
&...&...&207.5&$(5.1 \pm 0.5)\cdot 10^{3}$\\
\end{tabular}
\begin{tablenotes}
\item Note: The first column shows the samples analyzed and the ratios between the ice mixture components.  In the second, the temperature of the maximum desorption peak T$_{d}$ and the desorption energy E$_{d}$/K$_{b}$ of the ice mixtures condensed directly on the cold finger of the cryostat are reported. Third column is dedicated at the ice mixture desorption from olivine dust and here, the temperatures and desorption energies of the two observed peaks are reported. The desorption energies are given in kelvin, as E$_{d}$/K$_{b}$, where K$_{b}$ is Boltzmann's constant.
\end{tablenotes}
\end{threeparttable}
\label{tablefit}
\end{table*}

\begin{figure*}
\begin{center}
{\includegraphics[width=14cm]{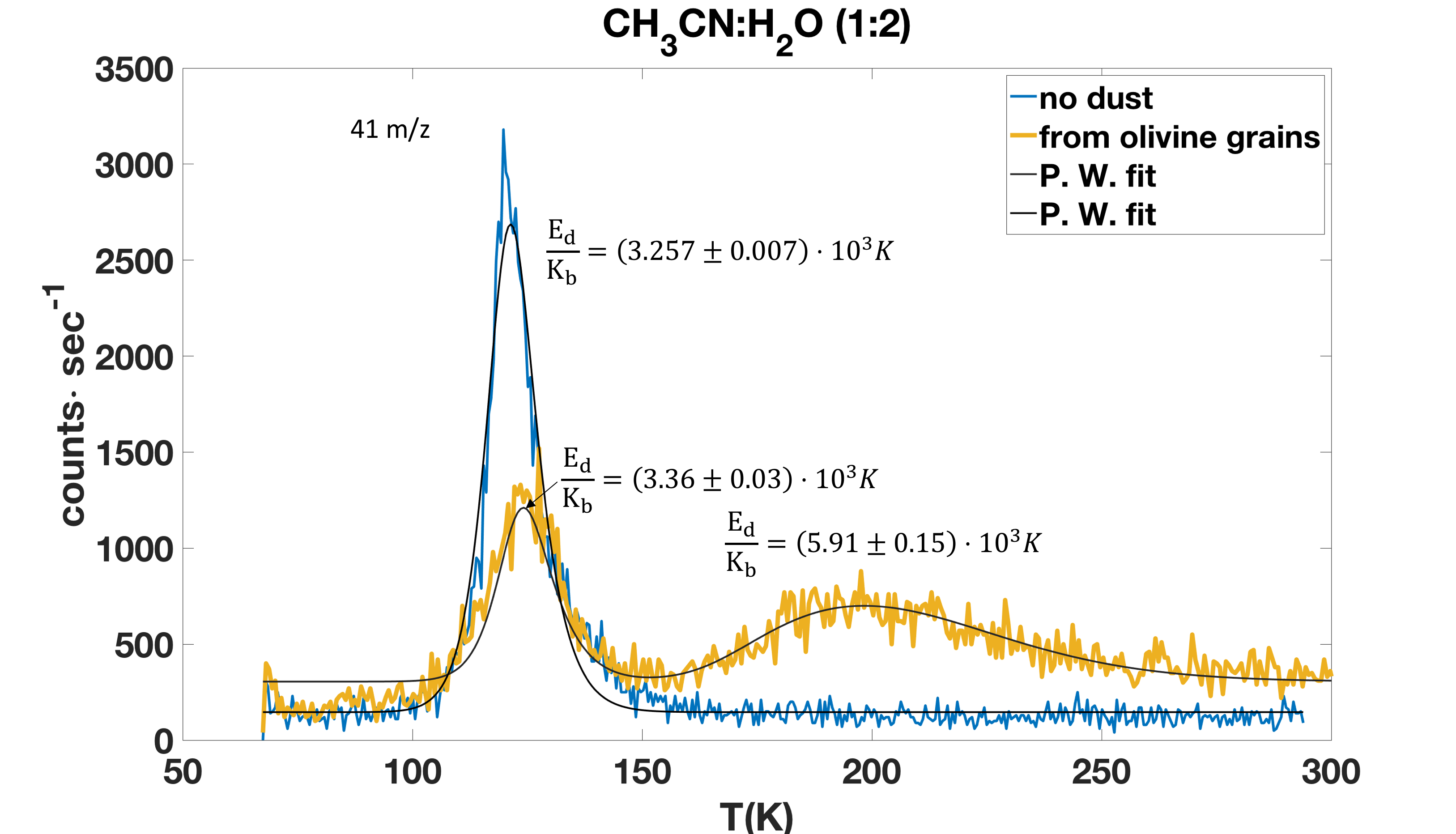}}
\caption{The figure shows two experiments. In both experiments we deposited in the UHV chamber an ice mixture CH$_{3}$CN:H$_{2}$O=1:2. \textbf{Blue curve}: CH$_{3}$CN desorption directly from the cold finger of cryostat cooled to 17 K. \textbf{Yellow curve}: CH$_{3}$CN desorption from olivine dust.
In presence of olivine, we found a first desorption peak lower in intensity than that found in the absence of grains and then we found a second wider desorption profile.
During the gas deposition, molecules once colliding to the cold grain surface can diffuse through the grains occupying more energetic favorable regions within the dust layer. During the TPD process such molecules are retained to higher temperatures before to be released to the gas phase.
\textbf{Black lines}: the TPD curves were fitted through the Polanyi - Wigner equation}\label{confrontoacetonitrilemiscelaconacqua}
\end{center}
\end{figure*}

\begin{figure*}
\begin{center}
{\includegraphics[width=14cm]{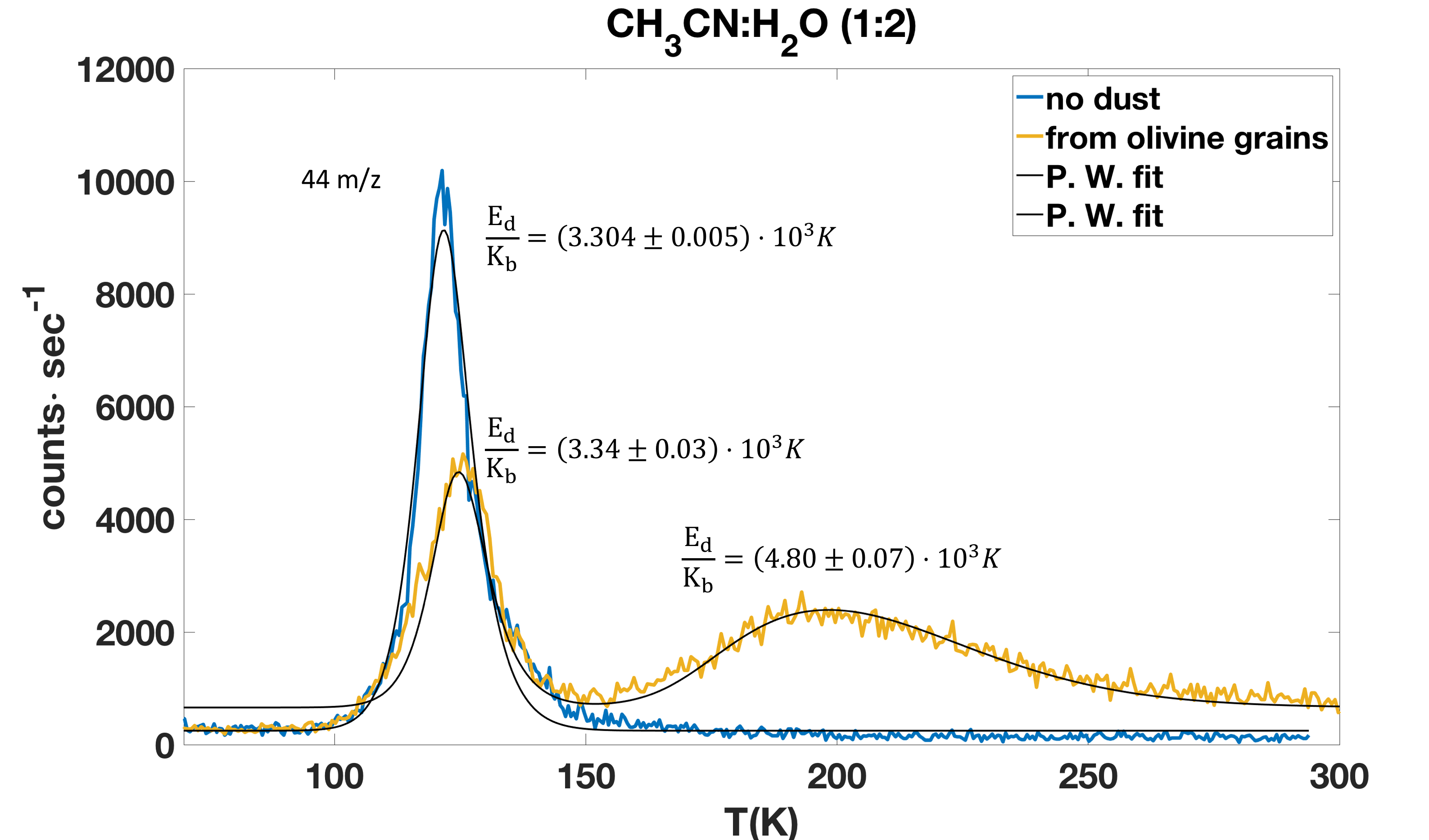}}
\caption{The figure shows two experiments. In both experiments we deposited in the UHV chamber an ice mixture CH$_{3}$CN:H$_{2}$O=1:2. \textbf{Blue curve}: m/z 44 desorption directly from the cold finger of cryostat cooled to 17 K. \textbf{Yellow curve}: m/z 44 desorption from olivine dust.
\textbf{Black lines}: the TPD curves were fitted through the Polanyi - Wigner equation}\label{confrontoacetonitrileacquaacet}
\end{center}
\end{figure*}

Figures \ref{confrontoacetonitrilemiscelaconacqua} and \ref{confrontoacetonitrileacquaacet} show the thermal desorption of the ice mixture CH$_{3}$CN:H$_{2}$O (1:2). 
Figure \ref{confrontoacetonitrilemiscelaconacqua} shows both the TPD curve of CH$_{3}$CN (signal at 41 m/z) desorbed from the smooth nickel-plated cold finger (blue curve) and that from olivine grains (yellow curve). 
In this experiment, first CH$_{3}$CN was mixed with water in the pre--chamber to obtain a 1:2 ratio and then deposited in the UHV chamber. Although acetonitrile was in a mixture with water, its TPD curves revealed similar results to those found when the molecule was deposited alone in the absence and presence of olivine. The blue curve (thermal desorption without grains) showed that the CH$_{3}$CN molecules started to desorb at 100 K reaching the maximum desorption peak at 121.4 K and at higher temperature the signal decreased and returned to the initial value already at 150 K. 
The yellow curve shows the thermal desorption from olivine dust. It shows that 50$\%$ of the deposited acetonitrile molecules desorbed at the same temperature found in absence of grains and 30$\%$ of molecules desorbed with a drift at 196.4 K and gradually dropping to the initial values. In presence of grains, 15$\%$ of molecules continued to desorb until 300 K, showing a gradual release of the molecules from the grains.
Figure \ref{confrontoacetonitrileacquaacet} shows the counts detected at 44 m/z during the desorption of the mixture from cold finger (blue curve) and from olivine grains (yellow curve). 
The blue curve showed only one desorption peak at 121.4 K. At this temperature, all molecules were already desorbed and at 150 K the signal already returned to the initial value. The yellow curve, that is the desorption from grains, exhibited a first peak at the same position as the desorption peak found in the absence of grains, but lower in intensity, and a shoulder at a higher temperature. In the absence of grains, all molecules of molecular weight 44 m/z desorbed around 120 K (signal intensity of $1\cdot 10^{4}$ counts$\cdot$sec$^{-1}$) . Instead, in the presence of grains the first peak was observed of intensity $5\cdot 10^{3}$ , so only 50$\%$ of the molecules desorbed from the grains. The second desorption peak was observed at 193.0 K.
The TPD curve at 44 m/z showed its desorption peak at the same temperature of acetonitrile and with higher peak intensity (the blue curves show an intensity of 3$\cdot$ 10$^{3}$ counts$\cdot$sec$^{-1}$ at 41 m/z and of 10$^{4}$ at 44 m/z). 
The simplest assignment for 44 m/z is CO$_{2}$, a common contaminant. However, we also carried out different background measurements in which we followed the signals detected by the mass spectrometer from 1 to 300 a.m.u as the temperature increased. From the background measurements, we found a signal at 44 m/z lower than 10$^{2}$ counts$\cdot$sec$^{-1}$, which is two orders of magnitude lower than the signal we observed during the mixture desorption. Even if we can not exclude that background CO$_{2}$ could contribute to the mass signal, its contribution can not justify the peak intensity observed.
A possible interpretations of this intense signal at 44 m/z is that it could be associated with the cracking in the mass spectrometer of dimethylamine CH$_{3}$-NH-CH$_{3}$ (45 m/z). In support of this interpretation, we observed a signal at 45 m/z with the same ratio compared to the signals at 41 and 44 m/z as found on the National Institute of Standards and Technology (NIST) Chemistry Webbook. The presence of dimethylamine could be a consequence of acetonitrile hydrogenation (e.g., \citealt{Nguyen2019}), but this interpretation will need to be further investigated and verified.

\begin{figure*}
\begin{center}
\subfigure[]
{\includegraphics[width=12cm]{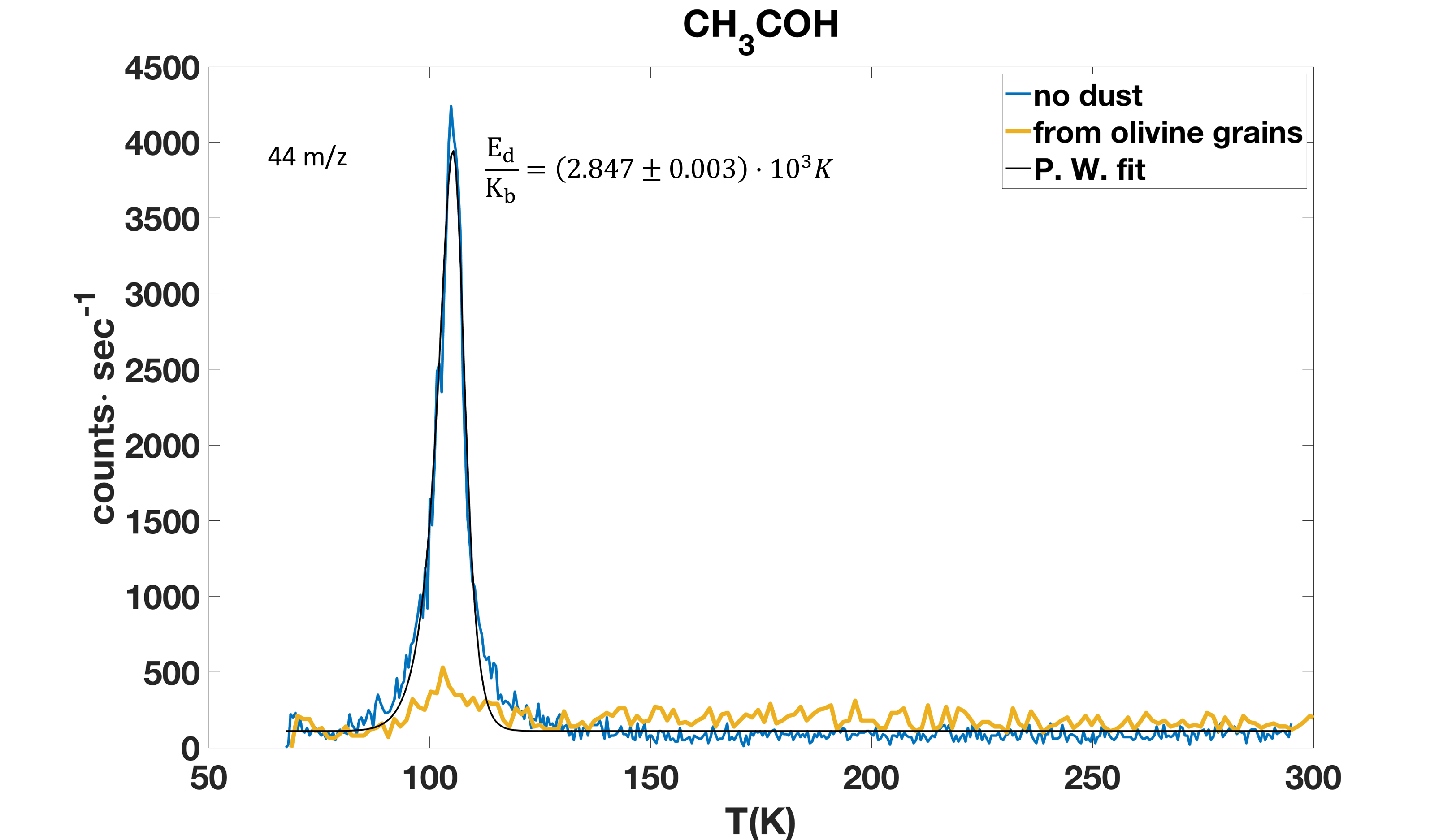}}
\subfigure[]
{\includegraphics[width=12cm]{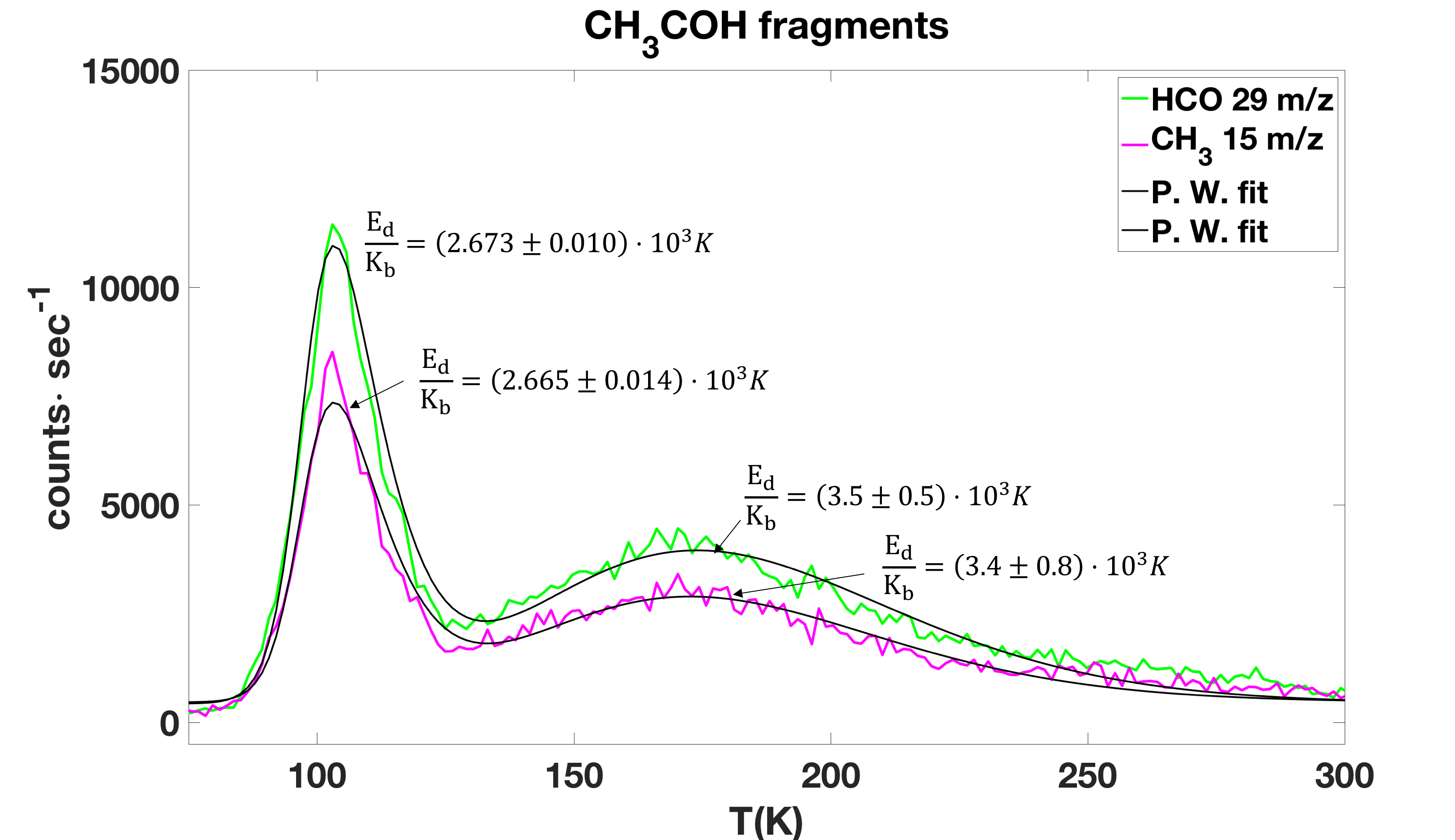}}
\caption{\textbf{(a)}The figure shows two experiments. In both experiments we deposited in the UHV chamber the same amount of acetaldehyde CH$_{3}$COH (partial pressure of CH$_{3}$COH in the pre-chamber 0.68 mbar). \textbf{Blue curve}: CH$_{3}$COH desorption directly from the cold finger of cryostat cooled to 17 K. \textbf{Yellow curve}: CH$_{3}$COH desorption from olivine dust. In presence of olivine, acetaldehyde dissociates almost completely into its two components: HCO and CH$_{3}$, shown in panel (b). \textbf{Black line}: the acetaldehyde TPD curve was fitted through the Polanyi - Wigner equation. \textbf{Figure (b)} shows the signal detected at 29 m/z, molecular weight of HCO (green curve) and the signal detected at 15 m/z, molecular weight of CH$_{3}$ (purple curve), while acetaldehyde desorbed from olivine dust. These TPD curves in presence of grains show a first desorption peak followed by a shoulder, fitted through the Polanyi - Wigner equation (\textbf{Black lines}).}\label{dissociazioneacetaldeide}
\end{center}
\end{figure*}

Figure \ref{dissociazioneacetaldeide}(a) exhibits two experiments: in both cases, the same amount of pure CH$_{3}$COH (partial pressure of 0.6 mbar) was deposited directly on the cold finger of the cryostat (blue curve) and on an olivine substrate (yellow curve) respectively. The blue curve shows us that as the temperature increased, the acetaldehyde molecules (signal at 44 m/z) already started to desorb around 90 K until reaching the desorption peak at 104.9 K (the counts detected by the mass spectrometer increased from $2\cdot 10^{2}$ to $4.24\cdot 10^{3}$ counts$\cdot$sec$^{-1}$, one order of magnitude greater). At this temperature all the acetaldehyde molecules deposited were already desorbed and at higher temperatures, the counts decreased until they returned to the initial value already around 130 K. So, a deposit of 0.6 mbar of CH$_{3}$COH on the cold finger of the cryostat in the ultra high vacuum regime provided a \textbf{44 m/z} signal of $4.24\cdot 10^{3}$ counts$\cdot$sec$^{-1}$ intensity at 104.9 K, showing that around 100 K all the molecules were already desorbed from the surface. On the other hand, when we deposited 0.6 mbar of CH$_{3}$COH on olivine dust (yellow curve), we found that the CH$_{3}$COH signal was significantly reduced. At the same temperature found without grains ($\sim$ 104 K), the yellow TPD curve revealed a slight increase in the intensity signal that ranged from $2\cdot 10^{2}$ to $5.3\cdot 10^{2}$ counts$\cdot$sec$^{-1}$, i.e. compared to the case without grains, only 12.5$\%$ of the molecules desorbed and passed in the gas phase. Figure \ref{dissociazioneacetaldeide}(b) shows the signal at 15 m/z (associated to the methyl radical CH$_{3}$) and 29 m/z (associated to the formyl radical HCO) detected by the mass spectrometer during the acetaldehyde desorption from olivine grains. 
These signals, observed at the same desorption temperature of acetaldehyde, are associated with the cracking patterns of acetaldehyde in the head of the mass spectrometer. From the acetaldehyde mass spectra available on NIST, the most intense signal occurs at 29 m/z followed by the 44 m/z with signal intensity of 80$\%$, and then by that at 15 m/z with 40$\%$. However, in the presence of grains in addition to a notable decrease in the 44 m/z signal intensity (only $\sim$500 counts$\cdot$sec$^{-1}$ compared to $4.24\cdot 10^{3}$ counts$\cdot$sec$^{-1}$ without grains), an increase in 29 and 15 m/z signal intensity was observed ($10^{4}$ counts$\cdot$sec$^{-1}$ and $8\cdot 10^{3}$ counts$\cdot$sec$^{-1}$ respectively). In particular, the 15 m/z signal was nearly as intense as the 29 m/z signal, much higher than the expected amount of the cracking patterns. This may suggest that the presence of olivine catalyses the carbon-carbon bond breaking in the acetaldehyde molecule. It was found in fact that carbon-carbon bond of adsorbed acetaldehyde breaks easily, occurring at low potential (e.g., \citealt{Stanley2008}).
Both signals showed a double desorption: a first peak at 103.0 K (the same temperature of sublimation of acetaldehyde) followed by a shoulder showing its maximum intensity increase at 170.2 K.

\begin{figure*}
\begin{center}
\includegraphics[width=14cm]{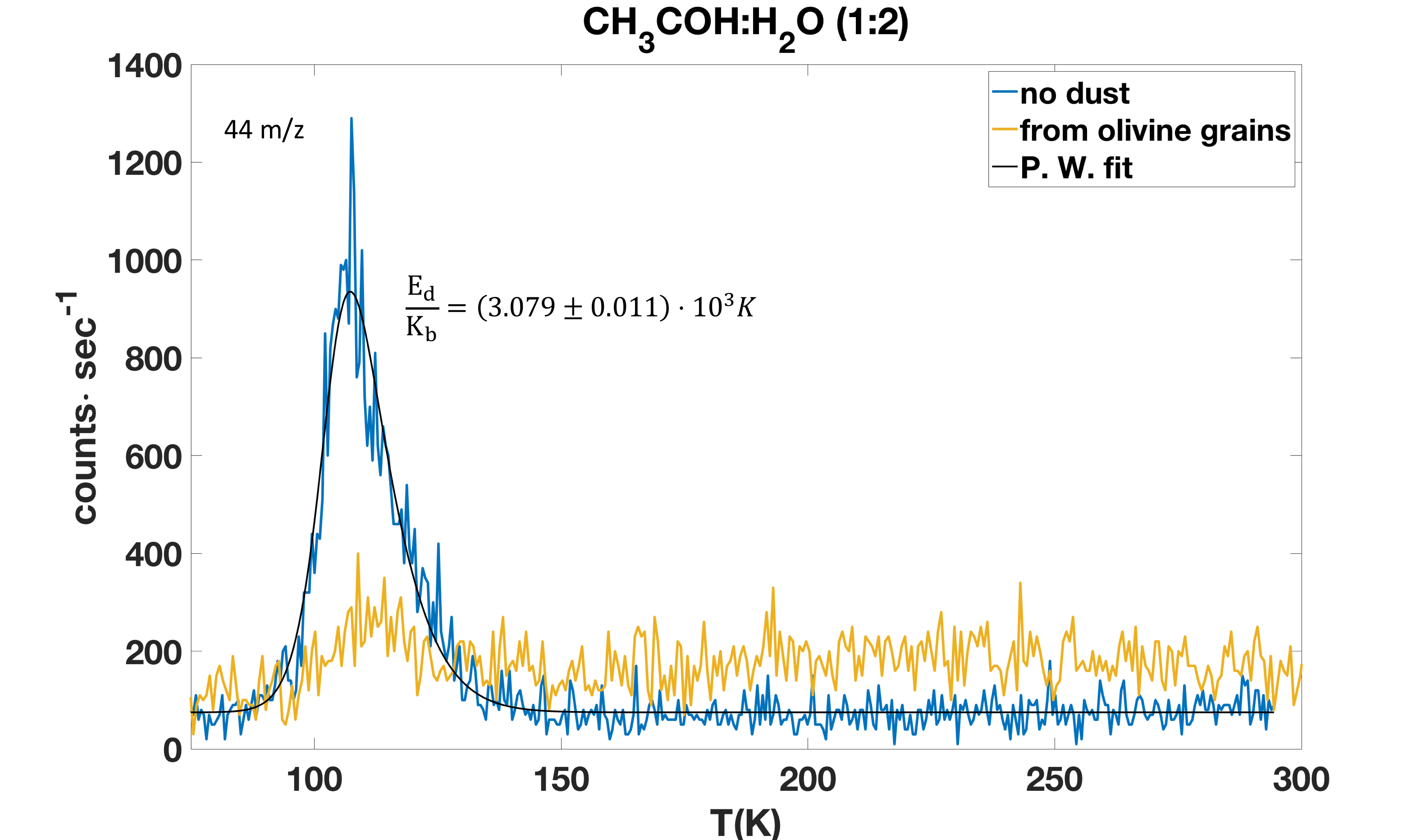}
\caption{The figure shows two experiments. In both experiments we deposited in the UHV chamber an ice mixture CH$_{3}$COH:H$_{2}$O=1:2. \textbf{Blue curve}: CH$_{3}$COH desorption directly from the cold finger of cryostat cooled to 17 K. \textbf{Yellow curve}: CH$_{3}$COH desorption from olivine dust.
In presence of olivine, acetaldehyde dissociates almost completely. \textbf{Black line}: the acetaldehyde TPD curve was fitted through the Polanyi - Wigner equation.}\label{confrontomiscelaacetaldeideacqua}
\end{center}
\end{figure*}

Figure \ref{confrontomiscelaacetaldeideacqua} presents the results obtained when acetaldehyde was deposited in the UHV chamber mixed with water. CH$_{3}$COH was first deposited with water in the pre--chamber and we realized a mixture 1:2 in favor of water through the partial pressures. Then, the mixture was deposited in the UHV chamber first directly on the cold finger of the cryostat (blue curve) and then on olivine dust (yellow curve). Although acetaldehyde was in a mixture with water, its TPD curves showed similar results to those found when the molecule was deposited alone (Fig. \ref{dissociazioneacetaldeide}). The blue curve (thermal desorption without grains) showed us that the CH$_{3}$COH molecules started to desorb at $\sim$ 90 K reaching the maximum desorption peak at 107.5 K and at higher temperature the signal decreased and returned to the initial value already at 130 K. When we deposited the same mixture on olivine dust (yellow curve), the 44 m/z intensity signal was significantly reduced and at the same temperature found without grains, we observed a slight increase in the intensity signal.

\begin{figure*}
\begin{center}
\subfigure[]
{\includegraphics[width=12cm]{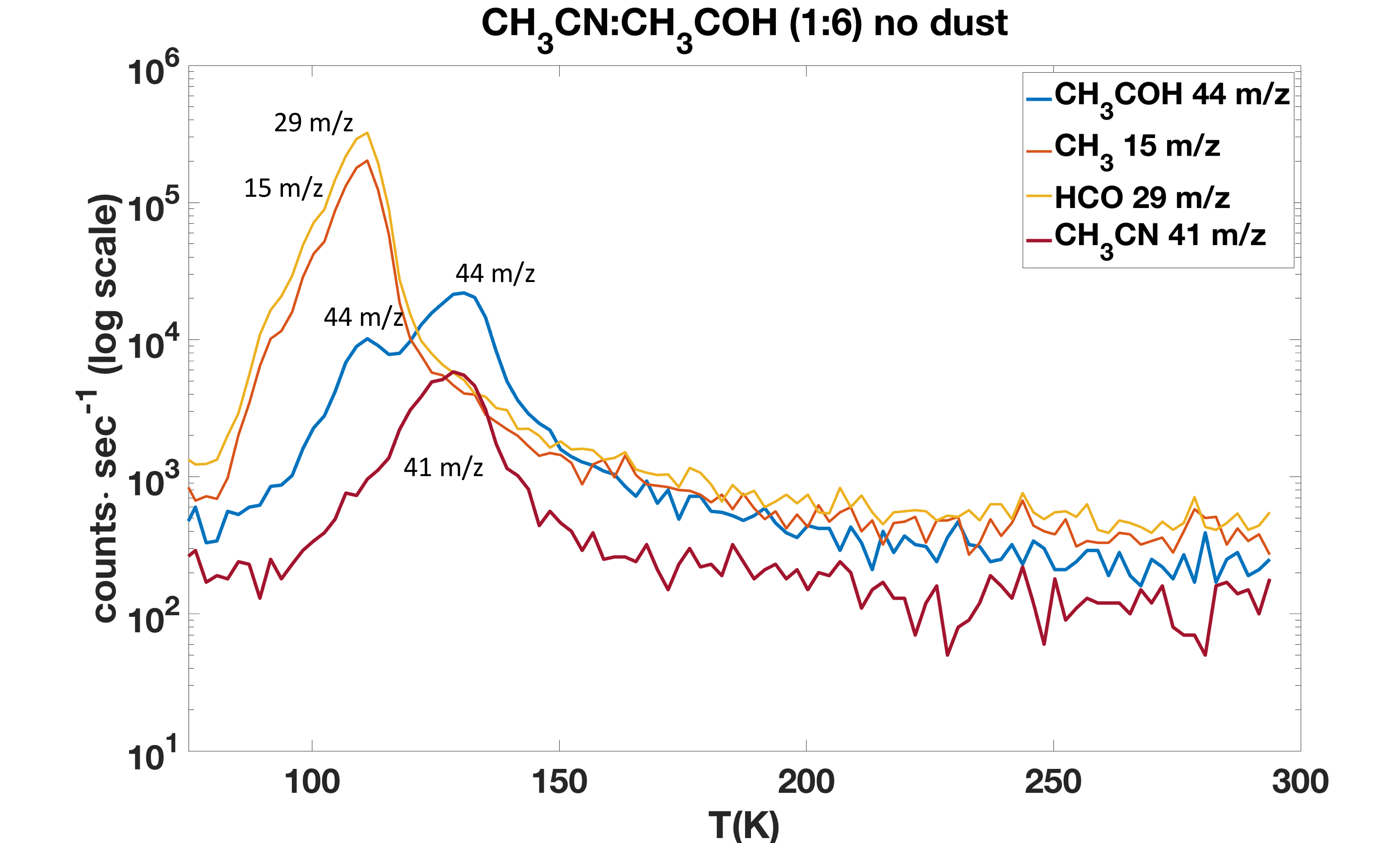}}
\vspace{5mm}
\subfigure[]
{\includegraphics[width=12cm]{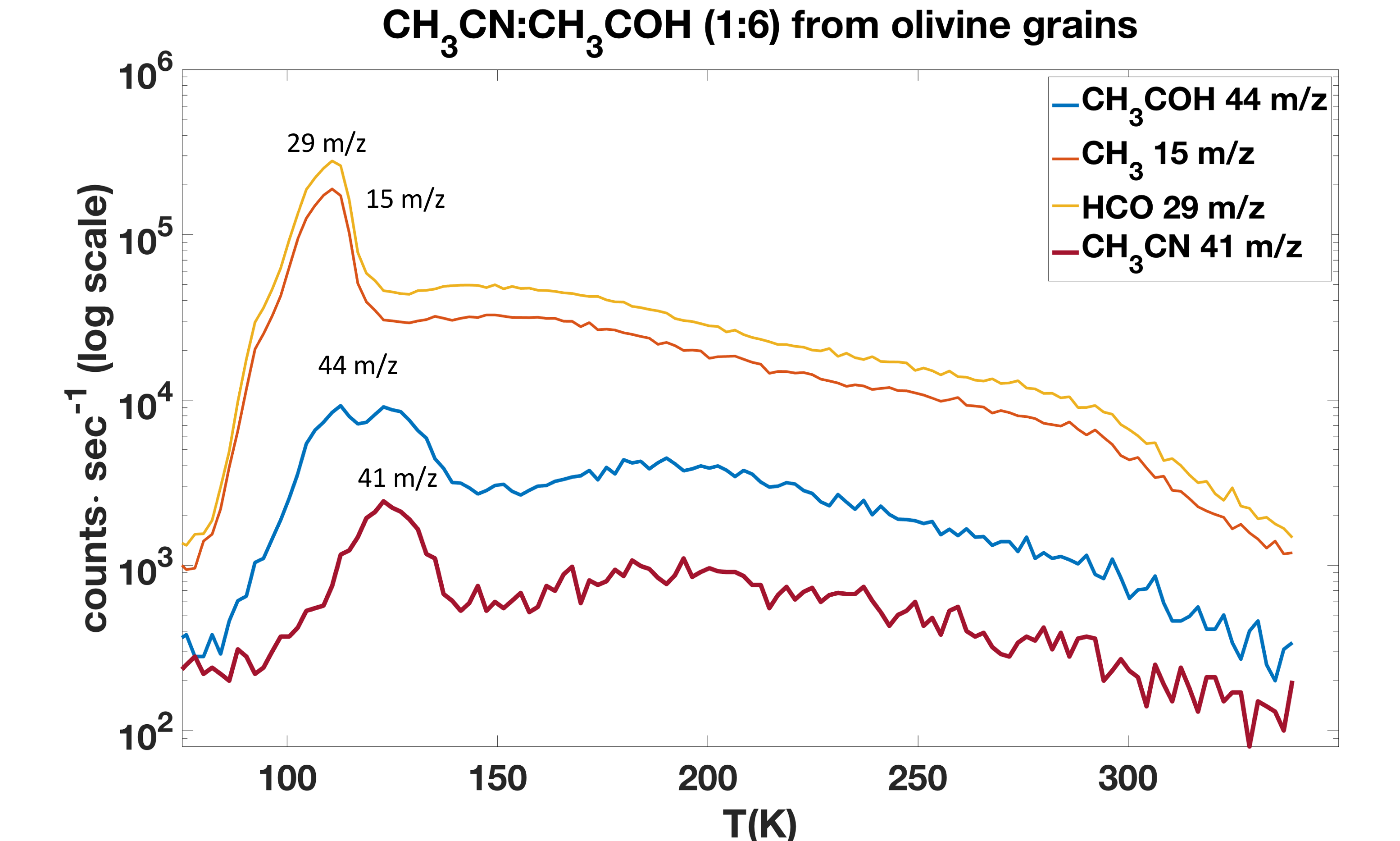}}
\caption{The figures show two experiments. In both experiments we deposited in the UHV chamber an ice mixture CH$_{3}$CN:CH$_{3}$COH=1:6. \textbf{Panel (a)}: ice mixture desorption directly from the cold finger of cryostat cooled to 17 K. \textbf{Panel (b)}: ice mixture desorption from olivine dust.
}\label{confrontomiscela1a6}
\end{center}
\end{figure*}

Through the experiments shown above, we found that acetaldehyde and its radicals desorbed around 100 K ($\sim$104 K when acetaldehyde was deposited alone Fig. \ref{dissociazioneacetaldeide}, $\sim$ 107 K mixed with water Fig. \ref{confrontomiscelaacetaldeideacqua}). In the acetonitrile case, the desorption peak was found at higher temperatures ($\sim$120 K) both when it was deposited alone or mixed with water (Figures \ref{confrontoacetonitrile} and \ref{confrontoacetonitrilemiscelaconacqua}). This difference in the desorption temperature between acetaldehyde and acetonitrile was also found when we deposited the two molecules together.
Figure \ref{confrontomiscela1a6} presents the desorption of the ice mixture CH$_{3}$CN:CH$_{3}$COH =1:6 from the cold finger of the cryostat (Fig. \ref{confrontomiscela1a6}(a)) and from the micrometric grains of olivine dust (Fig. \ref{confrontomiscela1a6}(b)) respectively. In Fig. \ref{confrontomiscela1a6}(a), we noted the acetaldehyde desorption peak, signal at 44 m/z, at 111.2 K. Above the acetaldehyde curve at the same temperature of sublimation, we found the TPD curves of its fragments: the methyl radical CH$_{3}$ (signal at 15 m/z) and formyl radical HCO (29 m/z). From the figure, we observed that the intensity of the radicals signal was an order of magnitude greater than that of acetaldehyde. 
At a higher temperature (130.7 K), we recognized the peak of acetonitrile (signal at 41 m/z). In correspondence with the desorption profile of acetonitrile, the signal at 44 m/z showed a new peak. The peak of the 44 m/z signal at 130 K is linked to acetonitrile. This is supported by the evidence that during the experiments with only acetonitrile and water, we observed a desorption peak corresponding to the 44 m/z signal at the same temperature of sublimation of acetonitrile and of greater intensity than the acetonitrile curve itself (Figs. \ref{confrontoacetonitrilemiscelaconacqua}  and \ref{confrontoacetonitrileacquaacet}).
In Figure \ref{confrontomiscela1a6}(a), the peaks at 130 K therefore describe the acetonitrile desorption (partial pressure of 0.3 mbar). The leftmost part of the graph instead describes the acetaldehyde desorption (partial pressure 1.8 mbar, six time higher than the acetonitrile amount). 
We also carried out this experiment in the presence of the olivine substrate (Fig. \ref{confrontomiscela1a6}(b)). In this case, all the TPD curves showed a shoulder when the temperature reached and exceeded the temperature of $\sim$ 200 K. In the presence of the grains, acetaldehyde molecules did not dissociate in favor of its radicals as in the experiments shown above. It must be considered that the amount of acetaldehyde deposited in this experiment was 6 time the acetonitrile (partial pressure of acetaldehyde 1.8 mbar), i.e. the amount of acetaldehyde deposited was greater than that deposited in the previous experiments.

\begin{figure*}
\begin{center}
\subfigure[]
{\includegraphics[width=12cm]{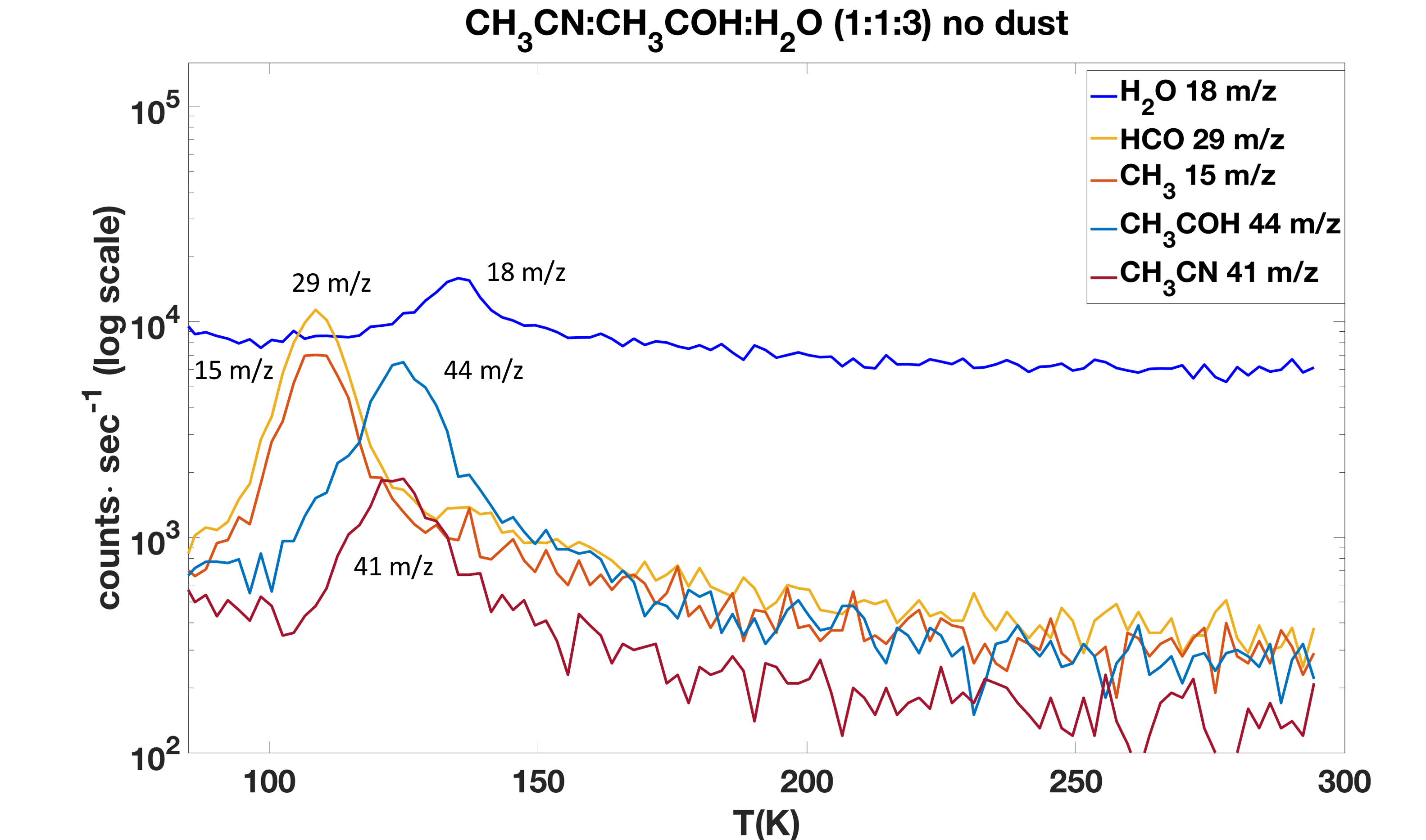}}
\vspace{5mm}
\subfigure[]
{\includegraphics[width=12cm]{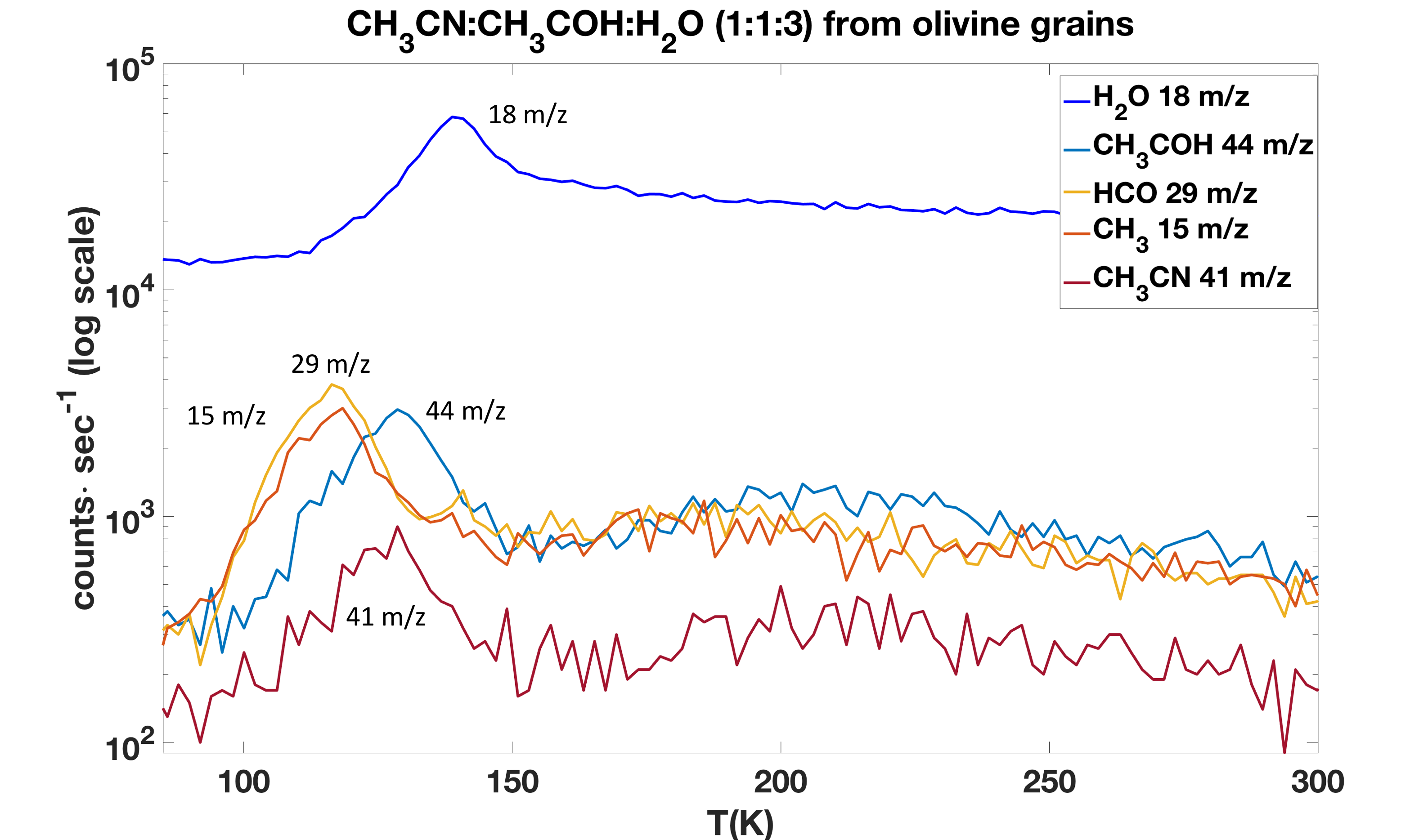}}
\caption{The figure shows two experiments. In both experiments we deposited in the UHV chamber an ice mixture CH$_{3}$CN:CH$_{3}$COH:H$_{2}$O=1:1:3. \textbf{Panel (a)}: ice mixture desorption directly from the cold finger of cryostat cooled to 17 K. \textbf{Panel (b)}: ice mixture desorption from olivine dust.}
\label{miscela3}
\end{center}
\end{figure*}

\begin{figure*}
\begin{center}
{\includegraphics[width=14cm]{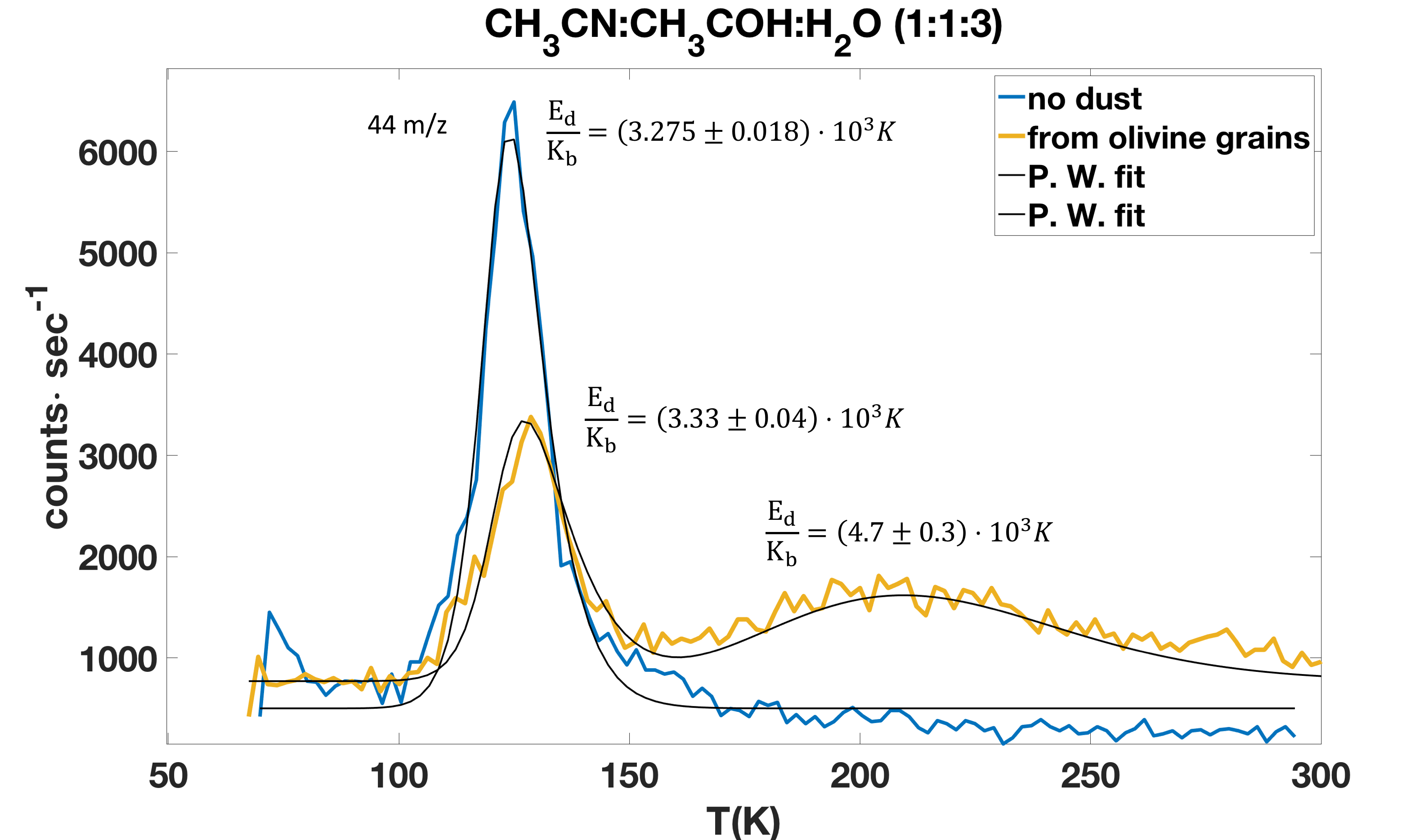}}
\caption{The figure shows two experiments. In both experiments we deposited in the UHV chamber an ice mixture CH$_{3}$CN:CH$_{3}$COH:H$_{2}$O=1:1:3. \textbf{Blue curve}: 44 m/z desorption directly from the cold finger of cryostat cooled to 17 K. \textbf{Yellow \textbf{curve}}: 44 m/z desorption from olivine dust.
In presence of olivine, we found a first desorption peak at the same temperature found in absence of grains but lower in intensity and then the curve shows a second wider desorption peak.
\textbf{Black lines}: the TPD curves were fitted through the Polanyi - Wigner equation}
\label{confrontomiscela3acetaldeide}
\end{center}
\end{figure*}

\begin{figure*}
\begin{center}
{\includegraphics[width=12cm]{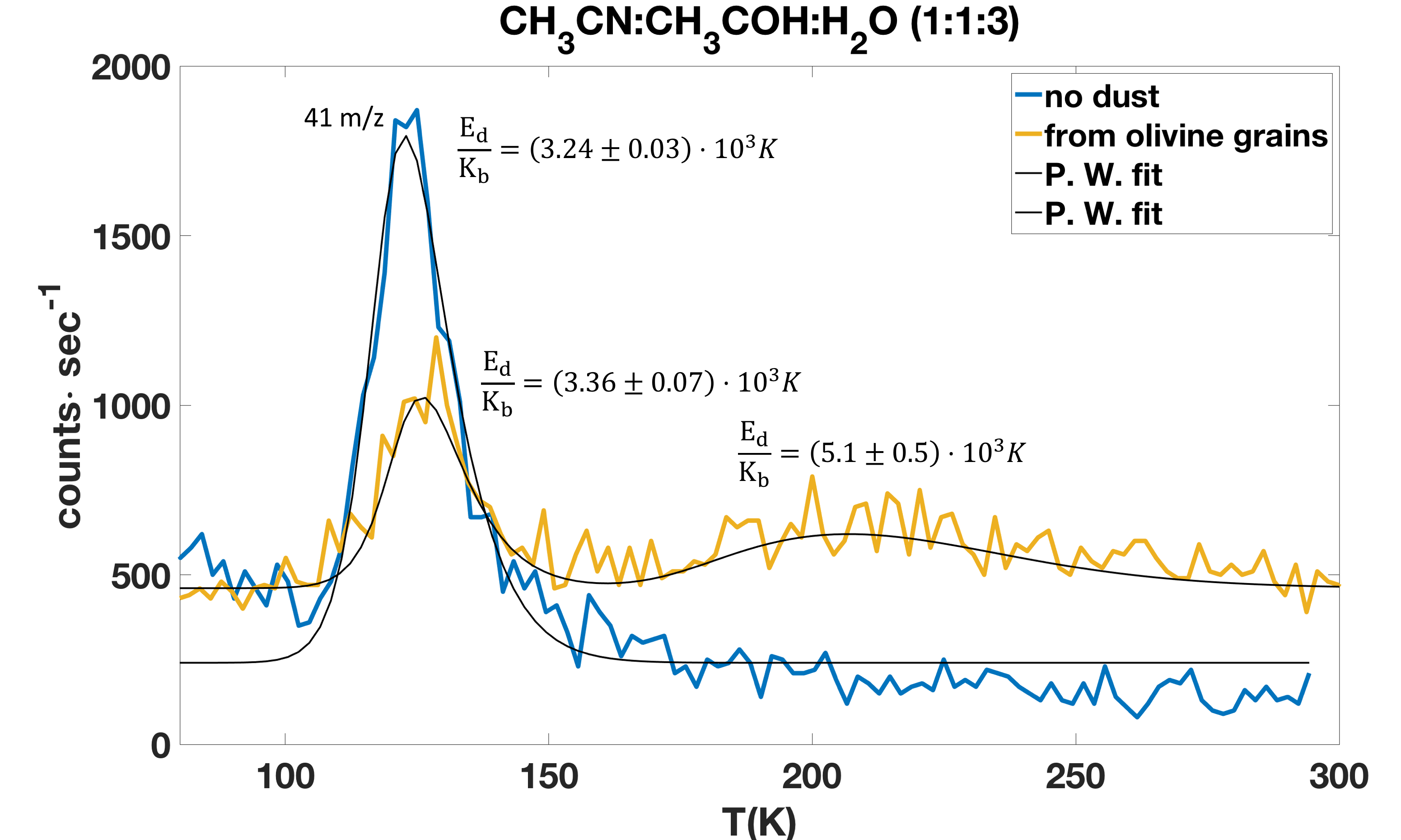}}
\caption{The figure shows two experiments. In both experiments we deposited in the UHV chamber an ice mixture CH$_{3}$CN:CH$_{3}$COH:H$_{2}$O=1:1:3. \textbf{Blue curve}: 41 m/z desorption directly from the cold finger of cryostat cooled to 17 K. \textbf{Yellow curve}: 41 m/z desorption from olivine dust.
\textbf{Black lines}: the TPD curves were fitted through the Polanyi - Wigner equation}
\label{confrontomiscela3acetonitrile}
\end{center}
\end{figure*}

Figure \ref{miscela3} displays the TPD of the ice mixture CH$_{3}$CN:CH$_{3}$COH:H$_{2}$O (1:1:3).
Figure \ref{miscela3}(a) shows the TPD curve of the ice mixture deposited directly on the cold finger of the cryostat. The desorption temperature at 108.7 K is associated with the acetaldehyde radicals formyl at 29 m/z (yellow curve) and methyl 15 m/z (orange curve). However, at this temperature, we did not observe the acetaldehyde peak at 44 m/z (light blue curve). This is due to acetaldehyde dissociation by the mass spectrometer in favor of its radicals. The experiments described above shown that acetonitrile desorbs at a higher temperature than acetaldehyde. Even in this case, we found CH$_{3}$CN peak (41 m/z) (red curve) at a higher temperature, that is 120.0 K. At this temperature a peak at 44 m/z (light blue curve) is observed.
H$_{2}$O (the highest blue curve) desorbed at 135.2 K an higher temperature than the other two molecules of the mixture.
Figure \ref{miscela3}(b) shows the TPD of the same mixture when deposited on olivine grains. What is noticed is that all the TPD curves showed a second peak at $\sim$ 200 K. In Figure \ref{confrontomiscela3acetaldeide} the desorption of the signal at 44 m/z from the cold finger (blue curve) and from olivine dust (yellow curve) is compared. The blue curve shows that the desorption started at $\sim$ 100 K reaching the maximum at 121.9 K (the signal intensity increased to $6.49\cdot 10^{3}$ counts$\cdot$sec$^{-1}$). At temperature higher than 121 K, the signal decreased and returned to the initial value at 150 K. The yellow curve shows that when we deposited the same mixture on olivine dust, 50$\%$ of the counts are measured at the same temperature previously found $\sim$ 123 K where the intensity of the yellow curve reached the value of $3.38\cdot 10^{3}$ counts$\cdot$sec$^{-1}$, and about 30$\%$ of the counts are observed at $\sim$ 200 K.
In the same way, the blue curve of the Figure \ref{confrontomiscela3acetonitrile} (CH$_{3}$CN desorption from the cold finger of the cryostat) showed only one desorption peak at 120.0 K: the signal started to desorb at 100 K, reached its maximum signal intensity at 120.0 K and then started immediately to decrease and returned to its initial value already at 150 K. The yellow curve instead (CH$_{3}$CN desorption from olivine dust) shows up that 60$\%$ of the acetonitrile molecules desorbed at the same temperature found in the absence of grains and the others molecules gradually desorbed as the temperature increased. Thus, during both 44 m/z and 41 m/z desorption from olivine experiments, we found a first desorption peak occurring at the same temperature than that found without grains but lower in intensity followed by a second desorption at higher temperature and hence, at higher desorption energy.\\
The Table \ref{tablefit} summarizes the best fit values ​​obtained through the Polanyi - Wigner equation for the desorption temperatures and energies both from cold finger and from olivine grains of pure and mixed molecules.

\section{Discussion and Astrophysical implication}
\subsection{Interaction with water}
A first result from our experiments concerns the desorption temperatures and energies found for CH$_{3}$CN and CH$_{3}$COH. Figures \ref{confrontoacetonitrile}, \ref{confrontoacetonitrilemiscelaconacqua}, \ref{dissociazioneacetaldeide}(a), and \ref{confrontomiscelaacetaldeideacqua} show the desorption from the cold finger (blue curves) and from olivine grains (yellow curves) both when the molecules of acetonitrile and acetaldehyde were deposited pure and mixed with water. Table \ref{tablefit} reports the values obtained by the Polanyi - Wigner equation for the desorption temperatures T$_{d}$ and energies E$_{d}$/K$_{b}$. As reported, we obtained similar results for T$_{d}$ and E$_{d}$/K$_{b}$ of acetonitrile both when it was deposited alone (Fig. \ref{confrontoacetonitrile}) and in mixture with water (Fig. \ref{confrontoacetonitrilemiscelaconacqua}), see Table \ref{tablefit}. Although, as we will see later, the shape of the TPD changes in the presence of the olivine substrate (yellow curve compared to the blue one), the position of the first desorption peak does not change. The T$_{d}$ and E$_{d}$/K$_{b}$ values ​​for the first peak are comparable even in the presence of grains, both in the experiments without the presence of water in the deposit phase and in the experiments with the water mixture. A similar comparison can be made for acetaldehyde, and once again, the results of T$_{d}$ and E$_{d}$/K$_{b}$ are comparable both in the experiment with the pure molecule and in a mixture with water (Figs. \ref{dissociazioneacetaldeide}(a), \ref{confrontomiscelaacetaldeideacqua}, and Table \ref{tablefit}).
 Moreover, we found that acetaldehyde and its radicals CH$_{3}$ and HCO desorb at about 100 K, while acetonitrile desorbs at slightly higher temperatures T$_{d}$ = 120 K. The difference in desorption temperature between acetaldehyde and acetonitrile was also found when we deposited the two molecules together (Fig. \ref{confrontomiscela1a6} (a) and (b)), and in mixture with water (Fig. \ref{miscela3} (a) and (b)). Many experimental works studied the release in the gaseous phase of gas mixtures and molecules trapped in amorphous water ice showing as the release of the most volatile component of the ice mixture occurs for discrete temperature ranges associated with changes in ice (e.g., \citealt{BarNun1988}, \citealt{Allamandola1988} and \citealt{Hudson1991}).
Although the ability of water to trap molecules with lower sublimation temperatures is thought to be one of the responsible processes for the release in the gaseous phase of the most volatile species contributing to the gas phase composition of the interstellar medium (e.g., \citealt{Aznar2010}, \citealt{Hassel2004}), in our experiments water does not appear to play such a role in the molecules desorption. The molecules and their radicals desorb at different T$_{d}$ characteristic of each molecular species regardless the presence of water in the deposit.  
However, we must take into account that in space the amount of water compared to other molecules is much greater than the 1:2 ratio used in this work. Through spectroscopic surveys, CH$_{3}$CN was detected in more than 10 comets and its molecular abundance relative to water is 0.01$\%$ \citep{Mumma2011}. In comet 67P/Churyumov-Gerasimenko, the deduced bulk abundances of the major volatile species, obtained with the ROSINA instrument on board the Rosetta orbiter, show CH$_{3}$CN/H$_{2}$O = 0.0059$\%$ (so a ratio $\sim$ 1:16000) and CH$_{3}$COH/H$_{2}$O=0.047$\%$ (1:2000) \citep{Rubin2019}. In comet Lovejoy, the amount of acetaldehyde is 0.1$\%$ compared to water \citep{ValBorro2018}, that is a 1:1000 ratio. In the laboratory, we could not use these ratios due to experimental needs. We were forced to maintain the total pressure in the pre-chamber below 2 mbar, to avoid the risk of compromising the mass spectrometer detector. At this pressure, for example, the amount of acetaldehyde needed to stay in a 1:1000 ratio is $10^{-3}$ mbar, a quantity too low to obtain a clear TPD curve. The fact that in interstellar ices the amount of H$_{2}$O compared to other species is much greater than the ratio used in the experiments may influence the role played by water in the desorption of other species. Another experimental limit is represented by grains. In the interstellar medium, the grains are micrometric fluffy amorphous silicate grains while in our measurements, we worked with a 100 $\mu$m dust layer thick made of crystalline olivine grains with sizes smaller than 5 $\mu$m. This aspect can change the conditions of interactions between water and the other molecules.
In our previous work \citep{Agnola2020}, we observed that water is responsible for the gaseous phase release of volatile species such as HCO and NH$_{2}$ obtained as photoproducts of formamide UV irradiation. To assess the role of UV irradiation and its link with water desorption, we plan to carry out a new series of TPD measurements of  the mixtures used in this paper irradiated with UV. 
\subsection{The role of grains}
The results obtained in this work show how the interactions between the molecules and the surface of the grains can drive the presence of iCOMs in the gaseous phase in environments where thermal heating has a dominant role. Comparing the TPD curves of pure water, acetonitrile and acetaldehyde from olivine grains (yellow curves), we notice that they show a similar shape: a first sharp desorption peak followed by a second wider peak at higher temperature. The first peak is associated to first-order desorption of each molecule interacting with the surface atoms of the mineral assemblage. The desorption energies of molecular adsorbates on olivine surface measured in this work indicate that the overall dipole moment of the molecules−solid surface complex formed on physisorption is dominated by the dipole moments of the individual molecules. Among the three molecules, acetaldehyde has the lowest dipole moment of 2.70 D to which corresponds the lowest desorption energy of 2847 K. Acetonitrile has a dipole moment of 3.92 D which corresponds desorption energy of 3130 K. Finally, water molecules have the highest dipole moment of 4.80 D to which corresponds the highest desorption energy of 3560 K. \\
However, the presence of two TPD peaks suggests that two different processes take place when molecules desorb from dust grains. In fact, all the TPD measurements from the smooth nickel-plated cold finger do not show the second peak. This means that the second peak is related to the presence of olivine grains. In our experiment, we deposited the molecular species on a layer about 100 $\mu$m thick made of olivine grains smaller than 5 $\mu$m in size. Molecules during the vacuum deposition process diffuse through the dust layer occupying all the available active adsorption sites on olivine grains in a random process. 
During the TPD process, the adsorbate, once thermally excited with kinetic energies approaching the desorption barrier, will be free to move on the surface until it thermalizes again due to the action of dissipation forces. At temperatures corresponding to the desorption energy or higher, the molecules begin to leave the adsorption sites and desorb. The molecules placed on the surface exposed to vacuum desorb producing the first TPD peak observed at lower temperature. This process is described by the Polanyi-Wigner equation. On the other hand, molecules adsorbed on the surfaces of grains placed deeper into the layer, once desorbed, will collide with the next grain surface by thermalizing again. Multiple hoping will be made before molecules are released to vacuum: molecules have to travel a longer path within the dust layer due to multiple desorption/adsorption processes on olivine grain surfaces thus, they are released to the gas phase with a delay. Hence, those diffusing molecules will yield the second wider desorption peak at higher temperature. Such diffusion phenomenon may become important in protoplanetary disks, where the submicron interstellar grains present in the parent molecular cloud start to accrete into hundreds of microns fluffy dust (e.g., \citealt{Testi2014}). Our experiments show that in protoplanetary regions with T$\geq$ 100 K, where we expect to no longer have acetonitrile and acetaldehyde in the solid phase because they are already desorbed, a fraction of these molecules can instead survive on fluffy grains and can be desorbed at temperatures of about 200 K. The ratio between the intensity of the two desorption peaks depends on the thickness of the dust layer and can provide hints on the fraction of molecules released at different temperatures in e.g., protoplanetary disks. We observed a ratio between the peaks of about 0.4, obtained by dividing the intensity of the second peak (maximum number in counts$\cdot$sec$^{-1}$) by the first most intense peak for each TPD curve of a given m/z.
Specifically, this ratio is 0.38 for the acetonitrile (Fig.\ref{confrontoacetonitrile}), and 0.36 for the radicals HCO and CH$_{3}$ (Fig.\ref{dissociazioneacetaldeide} (b)).
This means that porous grains of about 100 $\mu$m size, which are the result of coagulation of micrometer size grains, could retain 40$\%$ of the molecules at temperatures of about 200 K in contrast with the typical assumption that all molecules are desorbed at temperatures higher than 100 K. Hence, the diffusion by large porous grains is a valuable process which may enable the delivery of molecules to regions at high temperatures. \\
From astronomical observations, we know that protoplanetary disks are dynamic objects. The mass is transported continuously and accreted by the central star over a period of millions of years. It is essential to understand the effects of this evolution on the primitive materials it contains. Distribution of water and water-carried species in the disk undergo condensation, growth, transport, and vaporization (e.g., \citealt{Ciesla2006}). In addition to these processes, our TPD experiments show that volatiles ices rich in oxygen, nitrogen, and carbon can survive in the innermost part of the disk inside the snowlines of O-, N-, and S-bearing molecules. The diffusion is therefore a process enabling the delivery of water to Earth-like planets which form close to their star, ensuring the permanence of volatiles in the innermost part of the disk and shifting inward the position of the snowlines. \\
Another interesting aspect that was observed during the desorption from olivine grains is that the intensity of the signals of the two acetaldehyde radicals CH$_{3}$ and HCO increase compared to the cold finger desorption (Fig. \ref{dissociazioneacetaldeide} (a) and (b)) and compared to the amounts expected for cracking patterns in the mass spectrometer. Olivine could therefore accelerate the carbon-carbon bond breaking of acetaldehyde in favor of HCO and CH$_{3}$. If confirmed, this behavior may have implications in the astrophysical context where episodic warm-ups interspersed with a colder period can occur (for example a Fu Ori system between two different outbursts). In this condition, it could be possible that acetaldehyde is first desorbed from an icy grain, then adsorbed again in the colder region, and finally definitively desorbed to gas phase (new FU Ori outburst). From our experiment, we foresee that only a small fraction of acetaldehyde will desorb from the grains since it will be almost completely dissociated into HCO and CH$_{3}$ through multiple desorption-adsorption processes.

\subsection{Comparison with the observed abundance ratios}
CH$_{3}$COH and CH$_{3}$CN are detected at all the stages of the formation of a Sun-like star. It is interesting to compare the observed abundance ratios in the different evolutionary stages with our laboratory results. From our experiments, it is not immediate to find an abundance ratio between CH$_{3}$COH and CH$_{3}$CN, because we found that acetonitrile itself provides a contribution also to 44 m/z, the same molecular weight of acetaldehyde. If we look the Figure 9(b), we see that at 108.7 K only the acetaldehyde radicals are present, while at 120.0 K (T$_{d}$ of acetonitrile)  acetonitrile provides a signal to 44 m/z . From the ratio between the abundance of the 44 m/z signal at 120.0 K and the radical signals at 108.7 K, we can infer an abundance ratio CH$_{3}$COH/CH$_{3}$CN $\sim$1, indeed, we did not observe segregation of the molecules during the desorption process and the ratios present in the solid phase are kept in the gas phase. 
Therefore, in the case of thermal desorption, these laboratory studies tell us that the observed ratios between the molecules in the gas phase reflect the molecular composition of the grain surface.\\
From the observations, we know that in the protostellar shock L1157-B1 along the outflow driven by the low-mass Class 0 protostar L1157-mm, the CH$_{3}$COH/CH$_{3}$CN abundance ratio varies depending on the region of the shock: CH$_{3}$COH/CH$_{3}$CN is $\sim$1 in the region where previous observations indicate that the chemistry is dominated by the release of molecules from grains, instead, the abundance ratio is $\sim$ 0.1 in the region where gas-phase chemistry is likely to dominate (\citealt{Codella2009,Codella2015,Codella2017}, and \citealt{Podio2017}).
In hot-corinos around Class 0 protostars ($\sim$ 10$^{4}$ yr), the CH$_{3}$COH/CH$_{3}$CN abundance ratio ranges from $\sim$0.5 to 5 (e.g., \citealt{Belloche2020}), while the only available estimates for the disk around the young outbursting star V883 Ori indicate a CH$_{3}$COH/CH$_{3}$CN abundance ratio of $\sim$25 \citep{Lee2019}. Similarly to a hot corino, the molecules covering the icy grains of the disk are thermally desorbed following the outburst of the star. However, acetaldehyde is 25 times more abundant than acetonitrile, a much higher ratio than that found in protostellar objects, both in hot corinos where thermal desorption occurs and in the protostellar shocks where desorption is not thermal. This suggests that the molecular composition of the ices inherited by the early protostellar stages, once incorporated into the disk, may undergo a chemical reprocessing.

\section{Conclusions}
In this work, we investigated the desorption of ice mixtures from silicate olivine, simulating a process that realistically takes place in star-forming regions. 
We can summarize the found results in two fundamental points:
\begin{itemize}
\item The TPD experiments show how the interactions between the molecules and the surface
of the grains can drive the presence of molecules in the gaseous phase.
In the presence of grains, the TPD curves show a first sharp desorption peak at about 100 K and 120 K for acetaldehyde and acetonitrile respectively followed by a second wider peak at higher temperature and with about 40$\%$ intensity with respect to the first peak. So, 40$\%$ of the molecules are retained by fluffy grains of the order of 100 $\mu$m up to temperatures of 190 -210 K. This may be important in protoplanetary disks where the submicrometric interstellar grains begin to agglomerate into fluffy grains of hundreds of microns.
The diffusion of molecules on the silicate surface is a valuable process enabling the permanence of the ices in the inner part of the disk. 
This implies that O-rich and N-rich volatiles ice can survive up to $\sim$ 200 K, broading the snowlines of O- and N-bearing molecules, such as CH$_{3}$CN and CH$_{3}$COH. 
The presence of olivine can therefore determine the approach of the snowlines to the star and the presence of water and volatile species in Earth-like planets forming close to their star.
\item During the desorption process, we did not observe segregation of the molecules and so, the abundance ratios present in the solid phase are kept in the gas phase. 
This result implies that the observed abundance ratios in the gas phase reflect the molecular composition of the grain surface, if the observed molecules in gas phase are the result of thermal desorption process as in the case of the hot corinos or the disk around the young outbursting star V883 Ori.
The fact that the CH$_{3}$COH/CH$_{3}$CN ratio in the disk is $\sim$25 while in the hot corinos around Class 0 protostars ranges from $\sim$0.5 to 5 suggests that the molecular composition of ices inherited from the early protostellar stages  is reprocessed and undergoes formation and destruction processes.
\end{itemize}
In conclusion, through laboratory studies it is possible to improve our understanding of the chemical-physical interactions between molecules and the surface of grains, a process that can affect significantly the presence of molecular species both in the gas phase and in the small planets on short orbits. These studies offer the necessary support to the observational data and may help our understanding of the formation and origin of iCOMs providing  an estimate of the fraction of molecules released at various temperatures.
\paragraph{Acknowledgements}
This work was performed in the framework of M. A. Corazzi Ph.D. project.\\
The authors are grateful to the anonymous referees for very helpful suggestions. 
We wish to thank the Italian Space Agency for co-funding the Life in Space project (ASI N. 2019-3-U.0).\\
This work was supported by (i) the PRIN-INAF 2016 "The Cradle of Life - GENESIS-SKA (General Conditions in Early Planetary Systems for the rise of life with SKA)", and (ii) the program PRIN-MIUR 2015 STARS in the CAOS - Simulation Tools for Astrochemical Reactivity and Spectroscopy in the Cyberinfrastructure for Astrochemical Organic Species (2015F59J3R, MIUR Ministero dell'Istruzione, dell'Università della Ricerca e della Scuola Normale Superiore).\\
D.F. acknowledge the support of the Italian National Institute of Astrophysics (INAF) through the INAF Main Stream projects “ARIEL and the astrochemical link between circumstellar disks and planets” (CUP C54I19000700005) and “Protoplanetary disks seen through the eyes of new-generation instruments” (CUP C54I19000600005).
 \\

\bibliography{Bibliografia}

\begin{thebibliography}{}
\expandafter\ifx\csname natexlab\endcsname\relax\def\natexlab#1{#1}\fi

\bibitem[{\'Abrah\'am {et~al.}(2020)\'Abrah\'am, K\'osp\'al, Chen, \&
  Carmona}]{Abraham2020}
\'Abrah\'am, P., K\'osp\'al, A., Chen, L., \& Carmona, A. 2020, Proceedings of
  the International Astronomical Union, 345

\bibitem[{Andron {et~al.}(2018)Andron, Gratier, Majumdar, Vidal, Coutens,
  Loison, \& Wakelam}]{Andron2018}
Andron, I., Gratier, P., Majumdar, L., {et~al.} 2018, \mnras, 481, Issue 4

\bibitem[{\'Angel Satorre~Aznar \& Leliwa-Kopystynki(2010)}]{Aznar2010}
\'Angel Satorre~Aznar, M., \& Leliwa-Kopystynki, J. 2010, 38th COSPAR
  Scientific Assembly

\bibitem[{Arce {et~al.}(2008)Arce, Santiago-Garc\'ia, Jørgensen, Tafalla, \&
  Bachiller}]{Arce2008}
Arce, H.~G., Santiago-Garc\'ia, J., Jørgensen, J.~K., Tafalla, M., \&
  Bachiller, R. 2008, \apjl, 681, Issue 1

\bibitem[{Attard \& Barnes(1998)}]{attard1998}
Attard, G., \& Barnes, C. 1998, Oxford chemistry primers, 59, ALL

\bibitem[{Bachiller \& P\'erez~Guti\'errez(1997)}]{bachiller1997}
Bachiller, R., \& P\'erez~Guti\'errez, M. 1997, \apj, 487, Issue 1, L93

\bibitem[{Bacmann {et~al.}(2012)Bacmann, Taquet, Faure, Kahane, \&
  Ceccarelli}]{Bacmann2012}
Bacmann, A., Taquet, V., Faure, A., Kahane, C., \& Ceccarelli, C. 2012, \aap,
  541, id.L12

\bibitem[{Bar-Nun {et~al.}(1988)Bar-Nun, Kleinfeld, \& Kochavi}]{BarNun1988}
Bar-Nun, A., Kleinfeld, I., \& Kochavi, E. 1988, \prb, 38, Issue 11

\bibitem[{Belloche {et~al.}(2020)Belloche, Maury, Maret, Anderl, Bacmann,
  Andr\'e, Bontemps, Cabrit, Codella, Gaudel, Gueth, Lef\'evre, Lefloch, Podio,
  \& Testi}]{Belloche2020}
Belloche, A., Maury, A.~J., Maret, S., {et~al.} 2020, \aap, 635, id.A198

\bibitem[{Bennett {et~al.}(2005{\natexlab{a}})Bennett, Jamieson, Osamura, \&
  Kaiser}]{Bennett2005a}
Bennett, C.~J., Jamieson, C.~S., Osamura, Y., \& Kaiser, R.~I.
  2005{\natexlab{a}}, \apj, 624, Issue 2, 1097

\bibitem[{Bennett {et~al.}(2005{\natexlab{b}})Bennett, Osamura, Lebar, \&
  Kaiser}]{Bennett2005b}
Bennett, C.~J., Osamura, Y., Lebar, M.~D., \& Kaiser, R.~I. 2005{\natexlab{b}},
  \apj, 634, Issue 1, 698

\bibitem[{Bergner {et~al.}(2018)Bergner, Guzman, \"Oberg, Loomis, \&
  Pegues}]{Bergner2018}
Bergner, J.~B., Guzman, V.~G., \"Oberg, K.~I., Loomis, R.~A., \& Pegues, J.
  2018, \apj, 857, Issue 1

\bibitem[{Bianchi {et~al.}(2019)Bianchi, Codella, Ceccarelli, Vazart,
  Bachiller, Balucani, Bouvier, De~Simone, Enrique-Romero, Kahane, Lefloch,
  Lopez-Sepulcre, Ospina-Zamudio, Podio, \& Taquet}]{Bianchi2019}
Bianchi, E., Codella, C., Ceccarelli, C., {et~al.} 2019, \mnras, 483, Issue 2

\bibitem[{Biver \& Bockelee-Morvan(2019)}]{Biver2019}
Biver, N., \& Bockelee-Morvan, D. 2019, ACS Earth and Space Chemistry, 3, issue
  8

\bibitem[{Blake {et~al.}(1987)Blake, Sutton, Masson, \& Phillips}]{Blake1987}
Blake, G.~A., Sutton, E.~C., Masson, C.~R., \& Phillips, T.~G. 1987, \apj, 315

\bibitem[{Bockel\'ee-Morvan {et~al.}(2017)Bockel\'ee-Morvan, Rinaldi, Erard,
  Leyrat, Capaccioni, Drossart, Filacchione, Migliorini, Quirico, Mottola,
  Tozzi, Arnold, Biver, Combes, Crovisier, Longobardo, Blecka, \&
  Capria}]{BockeleeMorvan2017}
Bockel\'ee-Morvan, D., Rinaldi, G., Erard, S., {et~al.} 2017, \mnras, 469

\bibitem[{Bogelund {et~al.}(2019)Bogelund, McGuire, Hogerheijde, van Dishoeck,
  \& Ligterink}]{Bogelund2019}
Bogelund, E.~G., McGuire, B.~A., Hogerheijde, M.~R., van Dishoeck, E.~F., \&
  Ligterink, N. F.~W. 2019, \aap, 624, id.A82

\bibitem[{Boruah {et~al.}(2017)Boruah, Gogoi, Nath, \& Ahmed}]{Boruah2017}
Boruah, M.~J., Gogoi, A., Nath, B.~C., \& Ahmed, G.~A. 2017, \jqsrt, 196

\bibitem[{Brownlee \& Team(2006)}]{Brownlee2006}
Brownlee, D.~E., \& Team, S.~M. 2006, \baas, 38

\bibitem[{Caselli \& Ceccarelli(2012)}]{Caselli2012}
Caselli, P., \& Ceccarelli, C. 2012, \aap, Review 20

\bibitem[{Cazaux {et~al.}(2003)Cazaux, Tielens, Ceccarelli, Castets, Wakelam,
  Caux, Parise, \& Teyssier}]{Cazaux2003}
Cazaux, S., Tielens, A. G. G.~M., Ceccarelli, C., {et~al.} 2003, \apj, 593,
  Issue 1, L51

\bibitem[{Ceccarelli {et~al.}(2007)Ceccarelli, Caselli, Herbst, Tielens, \&
  Caux}]{Ceccarelli2007}
Ceccarelli, C., Caselli, P., Herbst, E., Tielens, A. G. G.~M., \& Caux, E.
  2007, Protostars and Planets V, University of Arizona Press, Tucson, 47

\bibitem[{Chaabouni {et~al.}(2018)Chaabouni, Diana, Nguyen, \&
  Dulieu}]{Chaabouni2017}
Chaabouni, H., Diana, S., Nguyen, T., \& Dulieu, F. 2018, \aap, 612, id.A47

\bibitem[{Chen {et~al.}(2000)Chen, Wang, \& He}]{Chen2000}
Chen, P.~S., Wang, X.~H., \& He, J.~H. 2000, \apss, 271

\bibitem[{Chuang {et~al.}(2020)Chuang, Fedoseev, Qasim, Ioppolo, Jager,
  Henning, Palumbo, van Dishoeck, \& Linnartz}]{Chuang2020}
Chuang, K.~J., Fedoseev, G., Qasim, D., {et~al.} 2020, \aap, 635, id.A199

\bibitem[{Ciesla \& Cuzzi(2006)}]{Ciesla2006}
Ciesla, F.~J., \& Cuzzi, J.~N. 2006, \icarus, 181, Issue 1, 178

\bibitem[{Cieza {et~al.}(2016)Cieza, Casassus, Tobin, Bos, Williams, Perez,
  Zhu, Caceres, Canovas, Dunham, Hales, Prieto, Principe, Schreiber,
  Ruiz-Rodriguez, \& Zurlo}]{Cieza2016}
Cieza, L.~A., Casassus, S., Tobin, J., {et~al.} 2016, Nature, 535, Issue 7611,
  258

\bibitem[{Codella {et~al.}(2015)Codella, Fontani, Ceccarelli, Podio, Viti,
  Bachiller, Benedettini, \& Lefloch}]{Codella2015}
Codella, C., Fontani, F., Ceccarelli, C., {et~al.} 2015, \mnras, 449, L11

\bibitem[{Codella {et~al.}(2009)Codella, Benedettini, Beltr\'an, Gueth, Viti,
  Bachiller, Tafalla, Cabrit, Fuente, \& Lefloch}]{Codella2009}
Codella, C., Benedettini, M., Beltr\'an, M.~T., {et~al.} 2009, \aap, 507, Issue
  2, L25

\bibitem[{Codella {et~al.}(2010)Codella, Lefloch, Ceccarelli, Cernicharo, Caux,
  Lorenzani, Viti, Hily-Blant, Parise, Maret, Nisini, Caselli, Cabrit, Pagani,
  Benedettini, Boogert, Gueth, Melnick, Neufeld, Pacheco, Salez, Schuster,
  Bacmann, Baudry, Bell, Bergin, Blake, Bottinelli, Castets, Comito, Coutens,
  Crimier, Dominik, Demyk, Encrenaz, Falgarone, Fuente, Gerin, Goldsmith,
  Helmich, Hennebelle, Henning, Herbst, Jacq, Kahane, Kama, Klotz, Langer, Lis,
  Lord, Pearson, Phillips, Saraceno, Schilke, Tielens, van~der Tak, van~der
  Wiel, Vastel, Wakelam, Walters, Wyrowski, Yorke, Borys, Delorme, Kramer,
  Larsson, Mehdi, Ossenkopf, \& Stutzki}]{Codella2010}
Codella, C., Lefloch, B., Ceccarelli, C., {et~al.} 2010, \aap, 518, id.L112

\bibitem[{Codella {et~al.}(2016)Codella, Ceccarelli, Cabrit, Gueth, Podio,
  Bachiller, Fontani, Gusdorf, Lefloch, Leurini, \& Tafalla}]{Codella2016}
Codella, C., Ceccarelli, C., Cabrit, S., {et~al.} 2016, \aap, 586, id.L3

\bibitem[{Codella {et~al.}(2017)Codella, Ceccarelli, Caselli, Balucani, Barone,
  Fontani, Lefloch, Podio, Viti, Feng, Bachiller, Bianchi, Dulieu,
  Jim\'enez-Serra, Holdship, Neri, Pineda, Pon, Sims, \&
  Spezzano}]{Codella2017}
Codella, C., Ceccarelli, C., Caselli, P., {et~al.} 2017, \aap, 605, id.L3

\bibitem[{Cohen \& Witteborn(1985)}]{Cohen1985}
Cohen, M., \& Witteborn, F.~C. 1985, \apj, 294, 345

\bibitem[{Collings {et~al.}(2004)Collings, Anderson, Chen, Dever, Viti,
  Williams, \& MCCoustra}]{Collings2004}
Collings, M.~P., Anderson, M.~A., Chen, R., {et~al.} 2004, \mnras, 354, 1133

\bibitem[{Corazzi {et~al.}(2020)Corazzi, Fedele, Poggiali, \&
  Brucato}]{Agnola2020}
Corazzi, M.~A., Fedele, D., Poggiali, G., \& Brucato, J.~R. 2020, \aap, 636,
  id. A63

\bibitem[{Cordiner {et~al.}(2015)Cordiner, Palmer, Nixon, Irwin, Teanby,
  Charnley, Mumma, Kisiel, Serigano, Kuan, Chuang, \& Wang}]{Cordiner2015}
Cordiner, M.~A., Palmer, M.~Y., Nixon, C.~A., {et~al.} 2015, Astrophysical
  journal letters, 800, Issue 1

\bibitem[{de~Val-Borro {et~al.}(2018)de~Val-Borro, Milam, Cordiner, Charnley,
  Coulson, Remijan, \& Villanueva}]{ValBorro2018}
de~Val-Borro, M., Milam, S.~N., Cordiner, M.~A., {et~al.} 2018, \mnras, 474,
  Issue 1

\bibitem[{Enrique-Romero {et~al.}(2016)Enrique-Romero, Rimola, Ceccarelli, \&
  Balucani}]{Enrique2016}
Enrique-Romero, J., Rimola, A., Ceccarelli, C., \& Balucani, N. 2016, \mnras,
  459, Issue 1

\bibitem[{Favre {et~al.}(2018)Favre, Fedele, Semenov, Parfenov, Codella,
  Ceccarelli, Bergin, Chapillon, Testi, Hersant, Lefloch, Fontani, Blake,
  Cleeves, Qi, Schwarz, \& Taquet}]{Favre2018}
Favre, C., Fedele, D., Semenov, D., {et~al.} 2018, \apjl, 862, Issue 1, article
  id. L2

\bibitem[{Fujiyoshi {et~al.}(2015)Fujiyoshi, Wright, \& Moore}]{Fujiyoshi2015}
Fujiyoshi, T., Wright, C.~M., \& Moore, T. J.~T. 2015, \mnras, 451, 3371

\bibitem[{Garrod \& Herbst(2006)}]{Garrod2006}
Garrod, R.~T., \& Herbst, E. 2006, \aap, 457, Issue 3

\bibitem[{Goesmann {et~al.}(2015)Goesmann, Rosenbauer, Bredeh\"oft, Cabane,
  Ehrenfreund, Gautier, Giri, Kr\"uger, Le~Roy, MacDermott, McKenna-Lawlor,
  Meierhenrich, Caro, Raulin, Roll, Steele, Steininger, Sternberg, Szopa,
  Thiemann, \& S.}]{Goesmann2015}
Goesmann, F., Rosenbauer, H., Bredeh\"oft, J.~H., {et~al.} 2015, Science, 349,
  Issue 6247

\bibitem[{Guzman {et~al.}(2018)Guzman, Guzman, Garay, Bronfman, \&
  Hechenleitner}]{Guzman2018}
Guzman, A.~E., Guzman, V.~V., Garay, G., Bronfman, L., \& Hechenleitner, F.
  2018, \apj Supplement Series, 236, Issue 2, article id. 45

\bibitem[{Hama {et~al.}(2011)Hama, Watanabe, Kouchi, \& Yokoyama}]{Hama2011}
Hama, T., Watanabe, N., Kouchi, A., \& Yokoyama, M. 2011, \apjl, 738, Issue 1,
  article id. L15

\bibitem[{Hassel(2004)}]{Hassel2004}
Hassel, G. 2004, \aas, 204, id.24.02 Bulletin of the American Astronomical
  Society, Vol. 36

\bibitem[{Henning(2010)}]{henning2010}
Henning, T. 2010, \araa, 48, 21

\bibitem[{Herbst \& van Dishoeck(2009)}]{Herbst2009}
Herbst, E., \& van Dishoeck, E.~F. 2009, \araa, 47, Issue 1, 427

\bibitem[{Honda {et~al.}(2003)Honda, Kataza, Okamoto, Miyata, Yamashita, Sako,
  Takubo, \& Onaka}]{Honda2003}
Honda, M., Kataza, H., Okamoto, Y.~K., {et~al.} 2003, \apj, 585, Issue 1

\bibitem[{Hudson \& Donn(1991)}]{Hudson1991}
Hudson, R.~L., \& Donn, B. 1991, \icarus, 94, Issue 2, 326

\bibitem[{Hudson {et~al.}(2008)Hudson, Moore, Dworkin, Martin, \&
  Pozun}]{Hudson2008}
Hudson, R.~L., Moore, M.~H., Dworkin, J.~P., Martin, M.~P., \& Pozun, Z.~D.
  2008, Astrobiology, 8, Issue 4, 771

\bibitem[{Iino {et~al.}(2020)Iino, Sagawa, \& Tsukagoshi}]{Iino2020}
Iino, T., Sagawa, H., \& Tsukagoshi, T. 2020, \apj, 890, Issue 2

\bibitem[{Kloprogge(2018)}]{Jacob2018}
Kloprogge, J. 2018, Spectroscopic Methods in the Study of Kaolin Minerals and
  Their Modifications, Springer Mineralogy. Springer, Cham

\bibitem[{Lai {et~al.}(2008)Lai, Kleyn, Rosca, \& Koper}]{Stanley2008}
Lai, S. C.~S., Kleyn, S. E.~F., Rosca, V., \& Koper, M. T.~M. 2008, \jcp, 112,
  48

\bibitem[{Lamberts {et~al.}(2019)Lamberts, Markmeyer, Kolb, \&
  Kastner}]{Lamberts2019}
Lamberts, T., Markmeyer, M.~N., Kolb, F.~J., \& Kastner, J. 2019, ACS Earth and
  Space Chemistry, 3, issue 6

\bibitem[{Le~Gal {et~al.}(2019)Le~Gal, Brady, Oberg, Roueff, \&
  Le~Petit}]{Gal2019}
Le~Gal, R., Brady, M.~T., Oberg, K.~I., Roueff, E., \& Le~Petit, F. 2019, \apj,
  886, Issue 2

\bibitem[{Lee {et~al.}(2019)Lee, Lee, Baek, Aikawa, Cieza, Yoon, Herczeg,
  Johnstone, \& Casassus}]{Lee2019}
Lee, J., Lee, S., Baek, G., {et~al.} 2019, Nature Astronomy, 3

\bibitem[{Lefloch {et~al.}(2017)Lefloch, Ceccarelli, Codella, Favre, Podio,
  Vastel, Viti, \& Bachiller}]{Lefloch2017}
Lefloch, B., Ceccarelli, C., Codella, C., {et~al.} 2017, \mnras, 469, Issue 1,
  L73

\bibitem[{Loomis {et~al.}(2018)Loomis, Cleeves, \"Oberg, Aikawa, Bergner,
  Furuya, Guzman, \& Walsh}]{Loomis2018}
Loomis, R.~A., Cleeves, L.~I., \"Oberg, K.~I., {et~al.} 2018, \apj, 859, Issue
  2

\bibitem[{Mart\'in-Dom\'enech {et~al.}(2020)Mart\'in-Dom\'enech, Oberg, \&
  Rajappan}]{MartinD2020}
Mart\'in-Dom\'enech, R., Oberg, K.~I., \& Rajappan, M. 2020, \apj, 894, Issue 2

\bibitem[{Mumma \& Charnley(2011)}]{Mumma2011}
Mumma, M.~J., \& Charnley, S.~B. 2011, \araa, 49, issue 1, 471

\bibitem[{Natta {et~al.}(2007)Natta, Testi, Calvet, Henning, Waters, \&
  Wilner}]{Natta2007}
Natta, A., Testi, L., Calvet, N., {et~al.} 2007, Protostars and Planets V, B.
  Reipurth, D. Jewitt, and K. Keil (eds.), University of Arizona Press, Tucson,
  951, 767

\bibitem[{Nguyen {et~al.}(2019)Nguyen, Fourr\'e, Favre, Barois, Congiu,
  Baouche, Guillemin, Ellinger, \& Dulieu}]{Nguyen2019}
Nguyen, T., Fourr\'e, I., Favre, C., {et~al.} 2019, \aap, 628, id.A15

\bibitem[{\"Oberg {et~al.}(2009)\"Oberg, Garrod, Van~Dishoeck, \&
  Linnartz}]{Oberg2009}
\"Oberg, K.~I., Garrod, R.~T., Van~Dishoeck, E.~F., \& Linnartz, H. 2009, \aap,
  504, 891

\bibitem[{\"Oberg {et~al.}(2015)\"Oberg, Guzman, Furuya, Qi, Aikawa, Andrews,
  Loomis, \& Wilner}]{Oberg2015}
\"Oberg, K.~I., Guzman, V.~V., Furuya, K., {et~al.} 2015, Nature, 520, Issue
  7546

\bibitem[{Ootsubo {et~al.}(2020)Ootsubo, Kawakita, Shinnaka, Watanabe, \&
  Honda}]{Ootsubo2020}
Ootsubo, T., Kawakita, H., Shinnaka, Y., Watanabe, J., \& Honda, M. 2020,
  \icarus, 338

\bibitem[{Picazzio {et~al.}(2019)Picazzio, Lukyanyk, Ivanova, Zubko, Cavichia,
  Videen, \& Andrievsky}]{Picazzio2019}
Picazzio, E., Lukyanyk, I.~V., Ivanova, O.~V., {et~al.} 2019, \icarus, 319

\bibitem[{Podio {et~al.}(2017)Podio, Codella, Lefloch, Balucani, Ceccarelli,
  Bachiller, Benedettini, Cernicharo, Faginas-Lago, Fontani, Gusdorf, \&
  Rosi}]{Podio2017}
Podio, L., Codella, C., Lefloch, B., {et~al.} 2017, \mnras, 470, Issue 1, L16

\bibitem[{Podio {et~al.}(2020)Podio, Garufi, Codella, Fedele, Bianchi,
  Bacciotti, Ceccarelli, Favre, Mercimek, Rygl, \& Testi}]{Podio2020a}
Podio, L., Garufi, A., Codella, C., {et~al.} 2020, \aap, 642, id.L7

\bibitem[{Poteet {et~al.}(2011)Poteet, Megeath, Watson, Calvet, Remming,
  McClure, Sargent, Fischer, Furlan, Allen, Bjorkman, Hartmann, Muzerolle,
  Tobin, \& Ali}]{Poteet2011}
Poteet, C.~A., Megeath, S.~T., Watson, D.~M., {et~al.} 2011, \apj, 733

\bibitem[{Przygodda {et~al.}(2003)Przygodda, van Boekel, \'Abrah\'am, Melnikov,
  Waters, \& Leinert}]{Przygodda2003}
Przygodda, F., van Boekel, R., \'Abrah\'am, P., {et~al.} 2003, \aap, 412, L43

\bibitem[{Rubin {et~al.}(2019)Rubin, Altwegg, Balsiger, Berthelier, Combi,
  De~Keyser, Drozdovskaya, Fiethe, Fuselier, Gasc, Gombosi, H\"anni, Hansen,
  Mall, Rème, Schroeder, Schuhmann, Sémon, Waite, \& Wampfler}]{Rubin2019}
Rubin, M., Altwegg, K., Balsiger, H., {et~al.} 2019, \mnras, 489, Issue 1, 594

\bibitem[{Sandford \& Allamandola(1988)}]{Allamandola1988}
Sandford, S.~A., \& Allamandola, L.~J. 1988, \icarus, 76, Issue 2, 201

\bibitem[{Shi {et~al.}(2015)Shi, Grieves, \& Orlando}]{Shi2015}
Shi, J., Grieves, G.~A., \& Orlando, T.~M. 2015, \apj, 804, Issue 1

\bibitem[{Shinnaka {et~al.}(2018)Shinnaka, Ootsubo, Kawakita, Yamaguchi, Honda,
  \& Watanabe}]{Shinnaka2018}
Shinnaka, Y., Ootsubo, T., Kawakita, H., {et~al.} 2018, \aj, 156, Issue 5,
  article id. 242

\bibitem[{Taniguchi {et~al.}(2020)Taniguchi, Guzman, Majumdar, Saito, \&
  Tokuda}]{Taniguchi2020}
Taniguchi, K., Guzman, A.~E., Majumdar, L., Saito, M., \& Tokuda, K. 2020,
  \apj, 898

\bibitem[{Taquet {et~al.}(2015)Taquet, L\'opez-Sepulcre, Ceccarelli, Neri,
  Kahane, \& Charnley}]{Taquet2015}
Taquet, V., L\'opez-Sepulcre, A., Ceccarelli, C., {et~al.} 2015, \apj, 804,
  Issue 2, article id. 81

\bibitem[{Testi {et~al.}(2014)Testi, Birnstiel, Ricci, Andrews, Blum,
  Carpenter, Dominik, Isella, Natta, Williams, \& Wilner}]{Testi2014}
Testi, L., Birnstiel, T., Ricci, L., {et~al.} 2014, Protostars and Planets VI,
  Henrik Beuther, Ralf S. Klessen, Cornelis P. Dullemond, and Thomas Henning
  (eds.), University of Arizona Press, Tucson, 914, 339

\bibitem[{Thelen {et~al.}(2019)Thelen, Nixon, Chanover, Cordiner, Molter,
  Teanby, Irwin, Serigano, \& Charnley}]{Thelen2019}
Thelen, A.~E., Nixon, C.~A., Chanover, N.~J., {et~al.} 2019, Icarus, 319

\bibitem[{van Boekel {et~al.}(2004)van Boekel, Min, Leinert, Waters, Richichi,
  Chesneau, Dominik, Jaffe, Dutrey, Graser, Henning, de~Jong, K\"ohler,
  de~Koter, Lopez, Malbet, Morel, Paresce, \& Preibisch}]{vanboekel}
van Boekel, R., Min, M., Leinert, C., {et~al.} 2004, \nat, 432, Issue 7016, 479

\bibitem[{van~'t Hoff {et~al.}(2018)van~'t Hoff, Tobin, Trapman, Harsono,
  Sheehan, Fischer, Megeath, \& van Dishoeck}]{vanthof2018}
van~'t Hoff, M. L.~R., Tobin, J.~J., Trapman, L., {et~al.} 2018, \apjl, 864,
  Issue 1, article id. L23

\bibitem[{Vastel {et~al.}(2014)Vastel, Ceccarelli, Lefloch, \&
  Bachiller}]{Vastel2014}
Vastel, C., Ceccarelli, C., Lefloch, B., \& Bachiller, R. 2014, \apjl, 795,
  Issue 1, article id. L2

\bibitem[{Walsh {et~al.}(2016)Walsh, Loomis, \"Oberg, Kama, van~'t Hoff,
  Millar, Aikawa, Herbst, Widicus~Weaver, \& Nomura}]{walsh2016}
Walsh, C., Loomis, R.~A., \"Oberg, K.~I., {et~al.} 2016, \apjl, 823, Issue 1,
  article id. L10

\bibitem[{Weinbruch {et~al.}(2000)Weinbruch, Palme, \& Spettel}]{Weinbruch2000}
Weinbruch, S., Palme, H., \& Spettel, B. 2000, \maps, 35, 161

\bibitem[{Willis {et~al.}(2020)Willis, Garrod, Belloche, Muller, Barger,
  Bonfand, \& Menten}]{Willis2020}
Willis, E.~R., Garrod, R.~T., Belloche, A., {et~al.} 2020, \aap, 636

\bibitem[{Woodney {et~al.}(2002)Woodney, A~Hearn, Schleicher, Farnham,
  McMullin, Wright, Veal, Snyder, de~Pater, Forster, Palmer, Kuan, Williams,
  Cheung, \& Smith}]{Woodney2002}
Woodney, L.~M., A~Hearn, M.~F., Schleicher, D.~G., {et~al.} 2002, Icarus, 157,
  Issue 1

\end{thebibliography}
\end{document}